# Harnessing Structural Disorder: Unraveling Hydrogen Evolution in Monolayer Amorphous Carbon via First-Principles Simulations and Machine-Learned Potentials

Sreehari M S, Ashutosh Krishna Amaram and Raghavan Ranganathan*

Department of Materials Engineering, Indian Institute of Technology Gandhinagar, Gandhinagar, 382355, Gujarat, India

*Corresponding author: rraghav@iitgn.ac.in

## Abstract

Disorder and defective coordination in the catalytic plane are crucial for enhancing the Hydrogen Evolution Reaction (HER) on two-dimensional catalysts. Amorphous materials are disordered, making them catalytically adaptive for many reactions. In this work, the HER capabilities of Monolayer Amorphous Carbon (MAC) were studied in comparison with crystalline carbon derivatives, such as pristine graphene (GE) and graphyne derivatives. MAC generated from melt-quench simulations revealed a diverse framework of predominantly $sp^2$ and $sp^3$ carbons with numerous 5-, 6-, and 7-membered rings. Density Functional Theory (DFT) calculations investigated free-energy variations in hydrogen adsorption for each material. According to Sabatier's principle, optimum activity is achieved when the Gibbs free energy ($\Delta G_H$) change approaches zero. Crystalline carbon materials possess limited active sites, with β-graphyne showing the best $\Delta G_H$ value of +0.34 eV. The adsorption study for MAC was conducted in 30 distinct local environments, where core structural properties were analyzed against varying radii. Calculations showed a $\Delta G_H$ distribution for MAC ranging from –0.02 eV to +1.35 eV. To evaluate activity across the entire MAC surface, a MACE MLIP foundation model was finetuned, achieving optimal energy and force fitting of 1.67 meV/atom and 29.15 meV/Å, respectively. The MLIP predicted $\Delta G_H$ values from −0.91 eV to +1.70 eV, with approximately 15% of sites exhibiting values below +0.25 eV. Feature analysis revealed that 7-membered rings, curvature, and ripple height enhance HER activity. Our findings suggest that, with careful optimization of local features, MAC can be tuned to compete with noble metal catalysts.

## 1. Introduction

The global transition toward carbon-free, clean, and sustainable energy has accelerated the search for new hydrogen production technologies. Hydrogen is an encouraging energy source of high energy density, which, upon combustion, yields zero carbon, making it invaluable for future energy systems [1],[2]. According to the International Energy Agency (IEA), hydrogen production reached 97 Mt in 2023, and among this production, only about 1% has low emissions. This minuscule fraction of sustainable hydrogen is produced by water electrolysis and steam reforming with efficient carbon capture. Electrolysers equipped with a suitable electrocatalyst split water into hydrogen ($H_2$) and oxygen ($O_2$), producing zero carbon dioxide. However, the production cost is currently 1.5 to 6 times higher than that of fossil fuel-based hydrogen production [3],[4]. This elevated cost is primarily due to the reliance on noble metal catalysts such as platinum (Pt) and iridium (Ir). A significant reduction in this cost can be achieved by discovering and developing more energy-efficient, low-cost catalysts for water splitting.

In recent years, there has been a rapid rise in the development of high-performance noble-metal-free 2D material catalysts for the hydrogen evolution reaction (HER) in water splitting [5],[6],[7],[8]. Two-dimensional carbon systems are unique in their electronic and structural properties; they have a high surface area, making them suitable for various energy applications. Graphene (GE) is a single atomic layer of $sp^2$ carbon atoms arranged in a hexagonal geometry, renowned for its mechanical, electric, and flexible nature. Another two-dimensional carbon allotrope class, Graphyne, is composed of sp- and $sp^2$-hybridized carbon atoms, forming a lattice of benzene rings connected by acetylene linkages [9]. Xun and colleagues theoretically investigated the electronic, optical, elastic, and thermal properties of α-, β-, and γ-graphyne (α-GY, β-GY, and γ-GY, respectively). They showed that their versatile geometries confer outstanding properties [10]. However, pristine crystalline carbon systems exhibit limited HER catalytic activity due to a lack of active sites; this can be enhanced through slight modifications, such as nitrogen doping, strain application, or vacancy creation [11],[12]. Anchoring single transition metals, such as Fe, Ni, and Pt, to graphene is another powerful strategy in which the metal atom acts as a catalytic centre [13],[14]. Recently, amorphous 2D materials have garnered widespread attention because of their disorder and defect-rich surfaces, which create favourable strain fields for catalytic and electronic activity [15], [16], [17]. Crystals possess a limited, well-defined set of reaction sites, while amorphous materials exhibit a broad distribution of possible sites. Unlike crystalline materials, which typically enable only surface-limited catalysis, amorphous catalysts can reconfigure under different conditions, allowing both volume and surface-confined electrocatalysis. They can convert themselves into catalytically active crystalline phases [18]. Numerous studies have shown improvements in catalytic activity using amorphous electrocatalysts derived from noble and earth-abundant metals [19], [20], [21]. For example, amorphous $PtSe_x$ ($1.2 < x < 1.3$) nanostructures possess fully activated, highly catalytic surfaces, achieving a current density nearly equivalent to that of a pure platinum surface [22]. Zhang et al. explored lithium-induced amorphization to enhance the HER performance of the $Pd_3P_2S_8$ crystals [23]. The amorphous NiMo alloy annealed at 400ºC attained an overpotential of 31 mV at 10 mA $cm^{-2}$ for the hydrogen evolution reaction [24]. In this study, we focus on *Monolayer Amorphous Carbon* (MAC), leveraging its diverse coordination environments to enhance the hydrogen evolution reaction (HER). Toh Chee-Tat et al. successfully demonstrated the generation of a free-standing, stable monolayer of amorphous carbon via laser-assisted chemical vapour deposition [25]. Furthermore, recent research has introduced slight modification to MAC, making it suitable for a wide range of applications [26]. We compare the hydrogen adsorption properties of MAC with various 2D carbon allotropes (e.g., graphene and graphynes) using plane-wave Density functional theory (DFT) and Machine-learned interatomic potential (MLIP).

MLIPs revolutionize atomistic simulations by combining first-principles accuracy with the efficiency needed for large-scale modeling. Universal MLIPs (U-MLIPs) show wide applicability across different atomic systems but typically need fine-tuning for best results on specific tasks [27], [28]. Among these, the MACE (Message-passing Atomic Cluster Expansion) framework stands out, with foundation models providing transferable starting points trained over vast chemical spaces [29]. Fine-tuning proves efficient for complex heterogeneous catalysis, where precise adsorption energy predictions are critical yet computationally prohibitive using conventional DFT [30], [31]. In fine-tuning, small set of high-fidelity DFT calculations is used to refine the pre-trained model for a specific catalytic system with needed weights and biases for energy and forces.

A representative local environment around the H adsorption site on MAC is isolated and analyzed via DFT to compute accurate $\Delta G_H$ values, fully breaking periodicity in this non-periodic cell setup. This approach yields near-exact results but is limited to a small number of sites due to high computational cost. To extend these isolated DFT studies and overcome wall-time limits, a machine learning interatomic potential (MLIP) is fine-tuned from an existing MACE foundation model using a small dataset. Additionally, the local structural features that drive favorable $\Delta G_H$ for H adsorption on MAC are investigated in a systematic manner.

## 2. Computational Details

### 2.1 Approach 1: DFT

The conventional melt-quench method was employed to model MAC using LAMMPS [32]. The ReaxFF force field was used to describe interatomic forces [33], [34]. The crystalline graphene of around 900 atoms was heated at 12000 K and equilibrated for 15 ps. In the computational setup, this high temperature is used to accelerate structural reorganization via thermal-driven amorphization. The system temperature was reduced to 2000 K in 100 ps, followed by an equilibration for 350 ps. Final quenching to 300 K was carried out over 200 ps, followed by 100 ps of equilibration. The structure relaxation and adsorption studies were carried out in the framework of Density functional theory using the Vienna ab initio simulation package (VASP) [35]. The core interactions were represented by Projector Augmented Wave (PAW) pseudopotential [36]. The generalized gradient approximation (GGA) with PBE exchange-correlation functional was used [37]. A plane-wave cutoff of 400 eV was used. We have used Gaussian smearing with a width of 0.05 eV to improve convergence. For Crystalline carbon derivatives like Graphene, α-GY, β-GY, and γ-GY, Monkhorst-Pack mesh of k-points 4×4×1, 2×2×1, 2×2×1, and 3×3×1, respectively. For an isolated MAC environment, a 1×1×1 k-points mesh was used. Grimme's DFT+D3 dispersion corrections were added for more accurate energy calculations [38]. The total energy convergence criterion in the self-consistent calculation and the ionic optimization convergence criterion was set to $10^{-5}$ eV and 0.05 eV/Å, respectively. A vacuum of 16 Å was applied in the z direction for each slab model to avoid periodic-image interactions. All the structural visualisations and modelling were done using OVITO[39] and ASE[40].

### 2.2 Approach 2: MACE MLIP for H Adsorption on MAC

#### 2.2.1 Dataset preparation

The datasets for MLIP train, test and validation were separately generated by combined MD and DFT simulations. To generate small set of diverse amorphous configurations, we have employed different MD melt-quench protocols on 200 atoms MAC, each protocol provided 30 different equilibrated MAC structures. The direct controlling factors like quench rate and final equilibration temperature for melt-quench were chosen 10 K/ps, 100 K/ps and 300 K, 400K, 500K respectively. To enhance void and local density diversity in the dataset, we have added three different vacancy concentrations of 3%, 5% and 6% above pristine structures. 24 possible combinations of quench rates, equilibration temperatures and vacancy concentrations (different protocols) composed the main dataset stream. To calculate energy and forces of each MD configuration, we performed static ab-initio DFT calculation on each. This way, we acquired diverse configurations of MAC along with its energy and forces. 12 random configurations from the entire MD outputs underwent DFT structure relaxation to capture near-equilibrium frames for accurate ground state energy estimation of MLIP. To capture relaxation pathways of hydrogen adsorption on MAC, we have sampled H adsorption on the 12 relaxed MAC configurations and conducted proper DFT relaxation until forces reduce to 0.01 eV/Å. From each relaxation trajectory, we extracted initial frames where energy differed from the prior frame by 0.08 eV, plus the final 5 converged frames to capture ground-state convergence accurately.

### 2.2.2 MACE foundation model and fine tuning

All the MLIP training was performed with MACE package [41]. We have employed naive fine tuning of the existing MACE foundation model MACE-matpes-pbe-omat-ft [42]. This foundation model was originally fine-tuned on MatPES (~400k structures from 281million MD snapshots, 16 billion environments; PBE level, 89 elements, no U-corrections), starting from MACE-OMAT-0 (OMAT dataset, PBE+U; excellent phonons). It combines OMAT-0 phonon accuracy with MatPES energy/force precision. Naive fine-tuning in MACE is the simplest fine-tuning protocol, restarting directly from a pre-trained foundation model's checkpoint and training only on new target dataset (smaller and specific), without replaying any original pre-training data. Atomic reference energies were determined self-consistently from the training data using the least-squares average method. Input features and atomic forces were normalized using RMS force scaling. Training was carried out for up to 1600 epochs with a batch size of 2, using the Adam optimizer and an initial learning rate of $1 \times 10^{-3}$. Exponential moving averaging (EMA) of model weights was applied with a decay factor of 0.99 to improve generalization and reduce overfitting. The combined loss function weighted energies and atomic forces with coefficients of 1.0 and 5.0, respectively. This is to prioritize accurate force reproduction, consistent with best practices for structure relaxation for adsorption study. All calculations were performed on a single NVIDIA A6000 GPU. A random 10% of the training data was held out as a validation set for monitoring convergence. The model with the lowest validation loss was selected as the final potential.

### 2.2.3 Structure Optimization using MACE MLIP and Local features extraction

The energy calculations and iterative relaxation using MACE MLIP was performed with MACECalculator module compiled within ASE (Atomic Simulation Environment) package. The model benchmarking was done with FIRE (Fast Inertial Relaxation Engine) [43] and hybrid FIRE+LBFGS (Limited-memory Broyden–Fletcher–Goldfarb–Shanno) [44] structure optimizers from ASE. Both optimizers used a final force convergence criterion of 0.01 eV/Å, although FIRE in the hybrid shifted to LBFGS after reaching a force value of 0.1 eV/Å. The structural features of H adsorbed site and its local environment were extracted using ASE and matscipy [45] packages. An imaginary spherical shell of 5Å radius was considered around the carbon atom attached to the hydrogen to represent the local environment.

### 2.3 Hydrogen adsorption study

Theoretical studies have shown that for an acidic medium, H adsorption is the key step in determining the HER performance [46]. The HER process can be either Volmer-Heyrovsky or Volmer-Tafel, as shown below.

Volmer step: $H^+ + e^- \rightarrow H^*$

Heyrovsky step: $H^* + H^+ + e^- \rightarrow H_2$

Tafel step: $2H^* \rightarrow H_2$

Therefore, we calculated the Gibbs free energy change for H adsorption at various possible adsorption sites for each catalyst to identify the most suitable one. The adsorption energy of hydrogen ($E_{ads}$) on a catalyst can be calculated from the following equation:

$$E_{ads} = E_{(surface-H)} - E_{surface} - \frac{1}{2}E_{H_2} \quad (1)$$

$E_{surface-H}$, $E_{surface}$, and $E_{H2}$ are the energies of the hydrogen adsorbed surface, the catalyst surface, and the hydrogen molecule, respectively. The thermodynamic variable, the Gibbs free energy change of adsorption, is:

$$\Delta G_H = E_{ads} + \Delta E_{ZPE} - T\Delta S \quad (2)$$

where $E_{ads}$ is the adsorption energy of the catalyst, $\Delta E_{ZPE}$ is the zero-point energy correction, T is the temperature, and ΔS is the entropy change. According to the pioneering work done by Nørskov [47], the exchange current density of the HER catalyst at pH = 0 can be derived from the descriptor $\Delta G_H$ as follows:

$$i_0 = -ek_0 \frac{1}{\left(1+exp^{\left(\frac{-\Delta G_H}{kT}\right)}\right)} \text{ for } \Delta G_H < 0 \quad (3)$$

$$\text{or } i_0 = -ek_0 \frac{1}{\left(1+exp^{\left(\frac{\Delta G_H}{kT}\right)}\right)} \text{ for } \Delta G_H > 0 \quad (4)$$

$k_0$ is the rate constant, and $k_B$ is the Boltzmann constant. The value of $k_0$ was set to 1.

## 3. Results and Discussions

### 3. 1 HER with Crystalline carbon derivatives

The detailed structural properties of graphene and graphynes are given in Table 1. Contextually, graphene has the densest planar atomic packing among 2D carbon, unlike the looser arrangement in α-GY. This is easily seen in Figure S1, which shows the void sizes for each structure. Bigger voids avoid the adsorption of small atoms like hydrogen, leading to a lower turnover rate. Due to its higher planar density, graphene exposes more atoms to any reaction than graphyne. However, the chemical environment of graphene is highly uniform and inert, which makes it less feasible for Hydrogen adsorption. Graphene has the fewest variety of possible H adsorption sites among them, due to its $sp^2$-only carbon configuration. Whereas α, β, and γ-GY have different concentrations of sp and sp2 hybridization carbons in their unit cells, as shown in Figure 1. We sampled H adsorption at possible sites for each variant (Figures S2-S6), and the best active sites are shown in the figure.

| **Derivative** | **a (Å)** | **b (Å)** | **Planar Density (Atoms/ Å$^2$)** | **sp-sp (Å)** | **sp-sp$^2$ (Å)** | **sp$^2$- sp$^2$ (Å)** |
|---|---|---|---|---|---|---|

| Graphene | 2.47 | 2.47 | 0.38 | - | - | 1.42 |
|---|---|---|---|---|---|---|
| α-GY | 6.97 | 6.97 | 0.19 | 1.23 | 1.40 | - |
| β-GY | 9.74 | 8.21 | 0.26 | 1.24 | 1.40 | 1.48 |
| γ-GY | 6.93 | 6.87 | 0.29 | 1.22 | 1.41 | 1.43 |

**Table 1**. Unit cell parameters and structural properties of graphene and graphynes

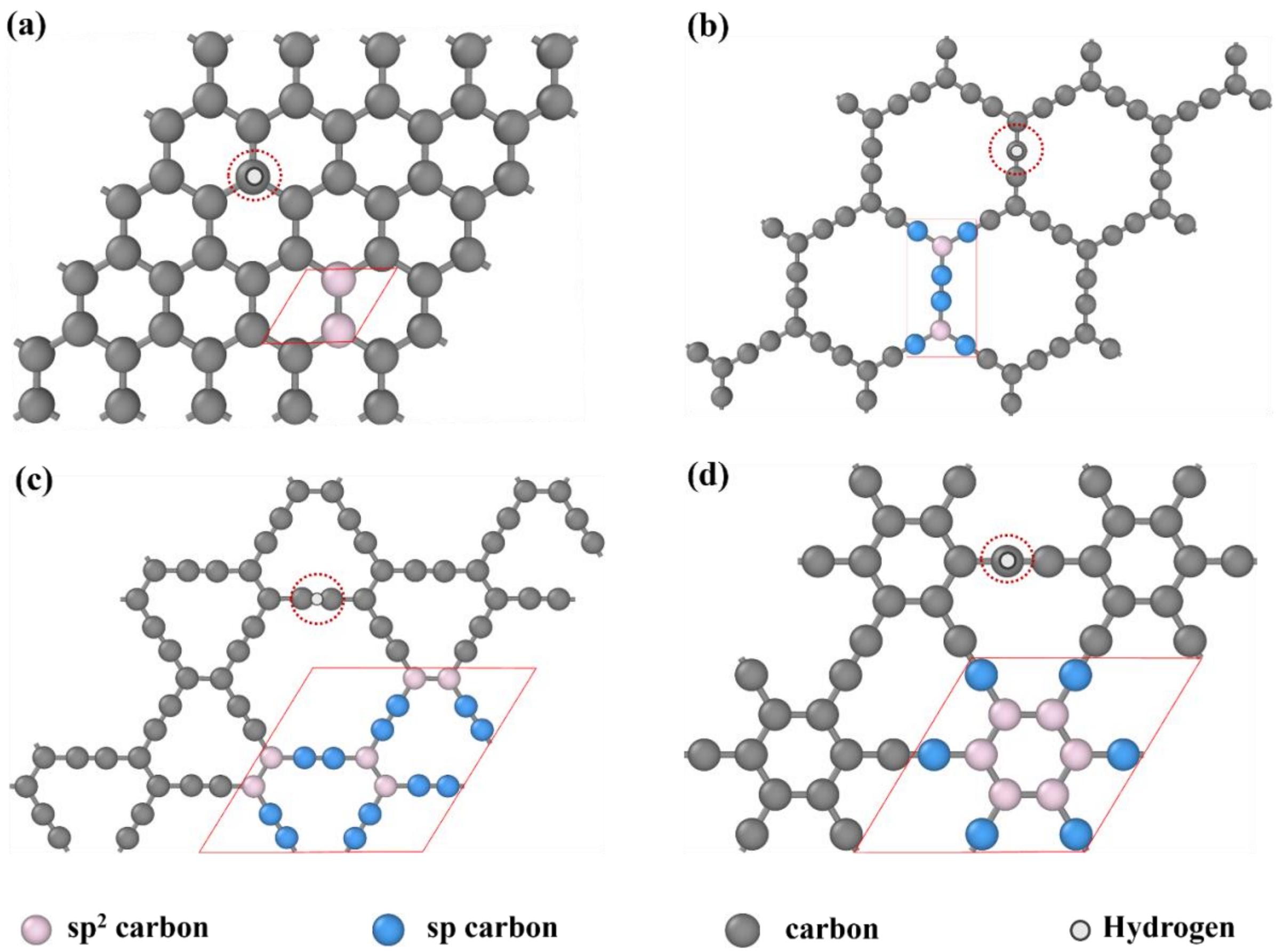

**Figure 1**: **(a)**, **(b)**, **(c)**, and **(d)** show the distribution of sp2 and sp carbon per unit cell in graphene, α-GY, β-GY, and γ-GY, respectively. The best hydrogen adsorption site for each species, ranked by $\Delta G_H$, is also shown in the same figure.

Among crystalline carbon structures, β-graphyne has the least positive $\Delta G_H$ value of +0.341 eV, indicating the best activity. According to Sabatier's principle, there will be a maximum rate of reaction when the interaction between reactants and catalysts is neither too strong nor too weak [48]. Thus, in theoretical studies, it has been proposed that the HER reaches its maximum reaction rate when the change in Gibbs free energy for hydrogen adsorption is close to zero [47]. α and γ-GY have a $\Delta G_H$ value of +0.538 and +0.668 eV, respectively, much lower than the +1.729 value of pristine graphene. Compared to β-GY, γ-GY possesses a higher planar density, making it a promising candidate for hydrogen adsorption. In addition to that, a study showed that γ-GY performs well in terms of structural stability and synthesis due to its higher acetylenic bonds per unit cell [49].

### 3.2 Modelling and structural analysis of the MAC catalyst

At first, a few random vacancies in a large-area monolayer graphene sheet consisting of ~900 atoms were introduced. This defective graphene underwent a melt-quench process, forming the planar MAC as shown in Figure 2 (a). On heating above the melting temperature, the crystalline arrangement undergoes randomization, forming a liquid-like state, followed by rapid quenching, during which the high-energy state suddenly arrests at a lower temperature. Here, quenching is conducted in two stages: the first with ultra-fast quenching to 2000K and the second with fast quenching to 300K. This is to reduce computational artifacts from extreme cooling rates in single-stage quenches and produce a stable amorphous configuration without unwanted crystallization. It also ensures uniform density and short-range order with varying ring sizes and a range of hybridisations mimicking experimental amorphous structures [50]. Figures 2(b) and 2 (c) show the ring statistics and hybridization details. Many 5- and 7-membered rings have been observed in MAC; however, 6-membered rings are predominantly observed, indicating localized crystallinity or graphene-like geometry. A minimal number of 3 and 4-membered rings are observed. 71.1 % of carbons in MAC are in the state of $sp^2$ hybridization, followed by 28.9% $sp^3$ state. Unlike crystalline derivatives, MAC has 0% of sp carbon because it lacks isolated linear geometries and instead forms localized, random networks. A spectroscopic study revealed that $sp^2$ and $sp^3$ bonding dominate over sp bonding in various amorphous carbons. Moreover, a monolayer enhances the $sp^3$ character but still does not exhibit sp bonding [51].

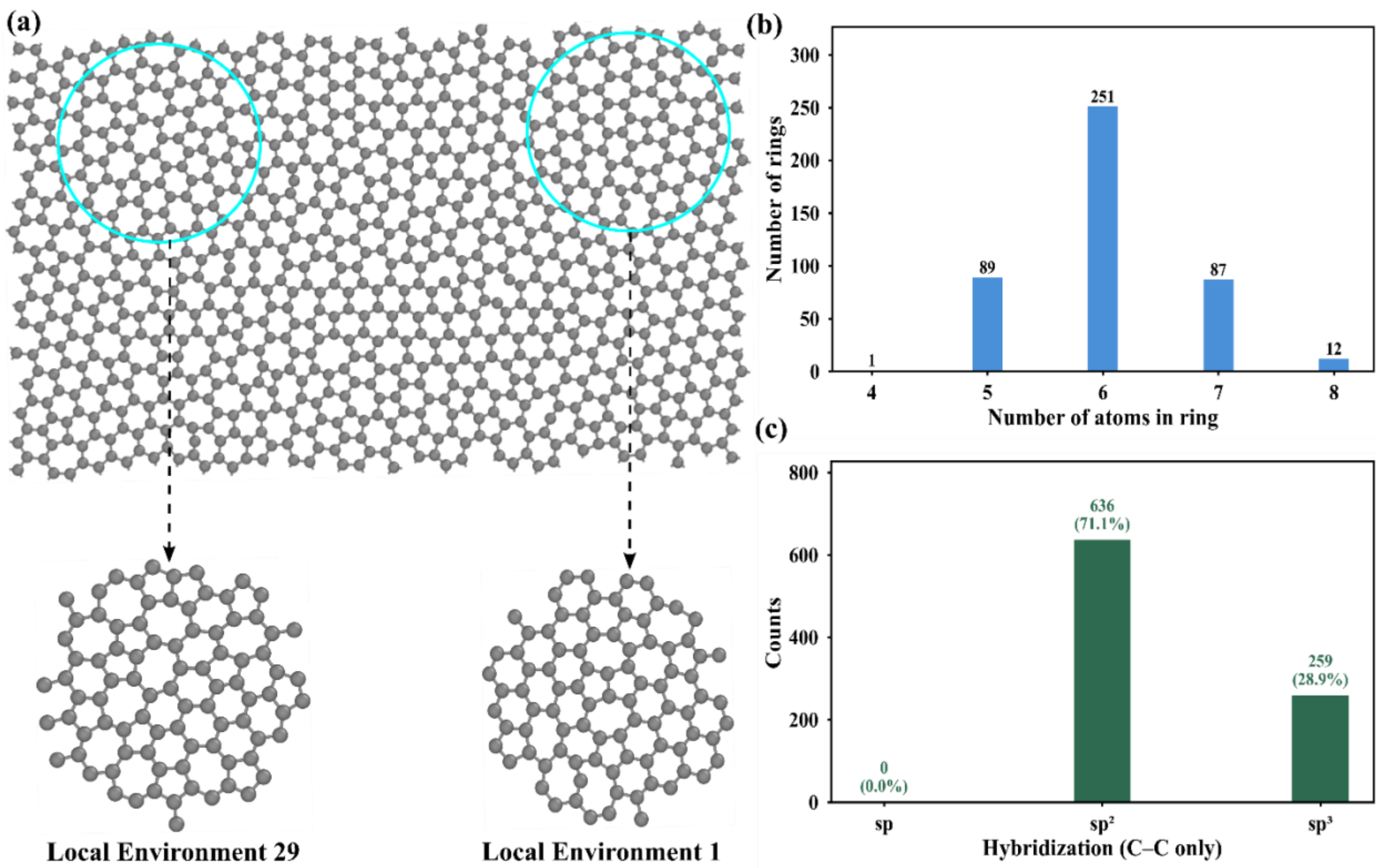

**Figure 2: (a)** The MAC structure obtained after melt-quench simulation. **(b)** Ring statistics and **(c)** Hybridization distribution in MAC.

### 3.3 HER performance of MAC

#### 3.3.1 Approach 1: DFT Evaluation of H Adsorption on MAC

For studying H adsorption, a single H atom was placed 2Å above the random target C atoms. The local environment within a specific radius around each sampled site was extracted and subjected to DFT structural relaxation in an isolated unit cell. A vacuum of 20 Å was added in all directions to avoid periodic interactions between the isolated environments. The periphery atoms were constrained in both the x and y directions to avoid in-plane movement during relaxation. Atomic interactions persisted even at infinite separation due to correlation effects. However, their effect diminished with increasing distance from the reference atom, particularly for basic electrostatic interactions that scale as *1/r* and become negligible beyond a certain radius. A convergence test was performed over different environment radii, ranging from 8.5 to 9.5 Å, to capture enough interactions affecting H adsorption. The local structural properties, specifically the core atom geometries, were studied for radii of 8.5, 9.0, and 9.5 Å. The details related to bond lengths and angles are shown in Table 2, all the values change very slightly after 9.0 Å. The core carbon atom charge magnitude remains unchanged at 0.04e beyond 9.0 Å. The radius was further optimized by H adsorption. The C-H bond length, the charge on the adsorbed H, and the Gibbs free energy change of H adsorption were calculated for all three radii. C-H bond length nearly remains constant across all the radii, while the Bader charge on adsorbed hydrogen and Gibbs free energy change reached convergence beyond 9.0 Å as shown in Figure 3.

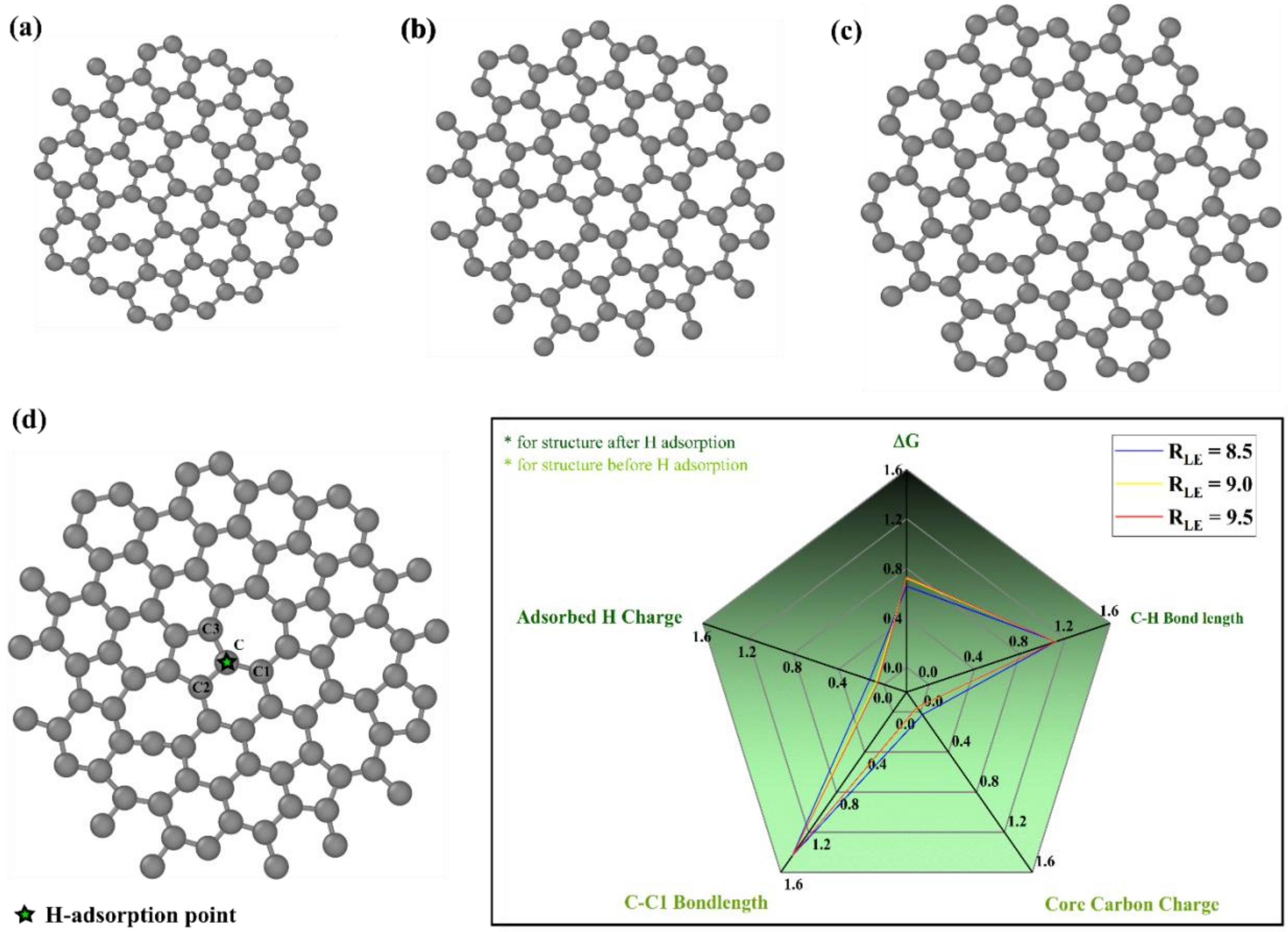

**Figure 3:** The local environment with radius **(a)** 8.5 Å, **(b)** 9.0 Å, and **(c)** 9.5 Å. **(d)** A radar plot showing the alignment of different properties with varying radius.

| Environment Radius (Å) | Bond length (Å) | | | Angle (degrees) | | | | | | | | |
|---|---|---|---|---|---|---|---|---|---|---|---|---|
| | | | | C1-C-C3 | | | C1-C-C2 | | | C2-C-C3 | | |
| | C-C1 | C-C2 | C-C3 | α | β | γ | α | β | γ | α | β | γ |
| 8.5 | 1.44 | 1.43 | 1.46 | 130.9 | 24.7 | 24.4 | 29.2 | 29.1 | 121.7 | 107.2 | 36.7 | 36.1 |
| 9 | 1.43 | 1.44 | 1.48 | 130.5 | 25.2 | 24.3 | 28.7 | 29.0 | 122.3 | 107.1 | 36.9 | 35.9 |
| 9.5 | 1.43 | 1.44 | 1.46 | 130.8 | 24.8 | 24.4 | 29.0 | 29.1 | 122.0 | 107.2 | 36.7 | 36.1 |

**Table 2:** The local structural properties of the extracted environment with different radii.

After optimizing the radius to 9.0 Å, we randomly selected 30 distinct local environments from the MAC surface and sampled hydrogen adsorption at the core carbon atom for each. The $\Delta G_H$ value for most of the environments came out positive, as with crystalline derivatives. One site (in local environment 28) showed an optimum value of –0.02 eV, followed by 0.04, 0.19, and 0.25 eV. The distribution of $\Delta G_H$ for the MAC surface is shown in Figure 4 (a). The variation of ring distribution and hybridization played a crucial role in deciding the H adsorption capacity of a particular site. The optimal local environment of MAC is compared with that of graphene and graphynes in terms of Gibbs free energy change, theoretical exchange-current values, and overpotential. The exchange current and overpotential corresponding to each $\Delta G_H$ value were calculated using equations 3 and 4, the plot is shown in Figure 4 (c) and Figure S7 respectively. These calculations highlighted the superiority of the MAC LE28 over other crystalline structures. If MAC can accommodate a higher number of such local environments, it could easily outperform existing crystalline carbon catalysts. This way, a robust mapping from amorphous synthesis to planar diversity would enable targeted amorphous catalysts for specific catalytic applications.

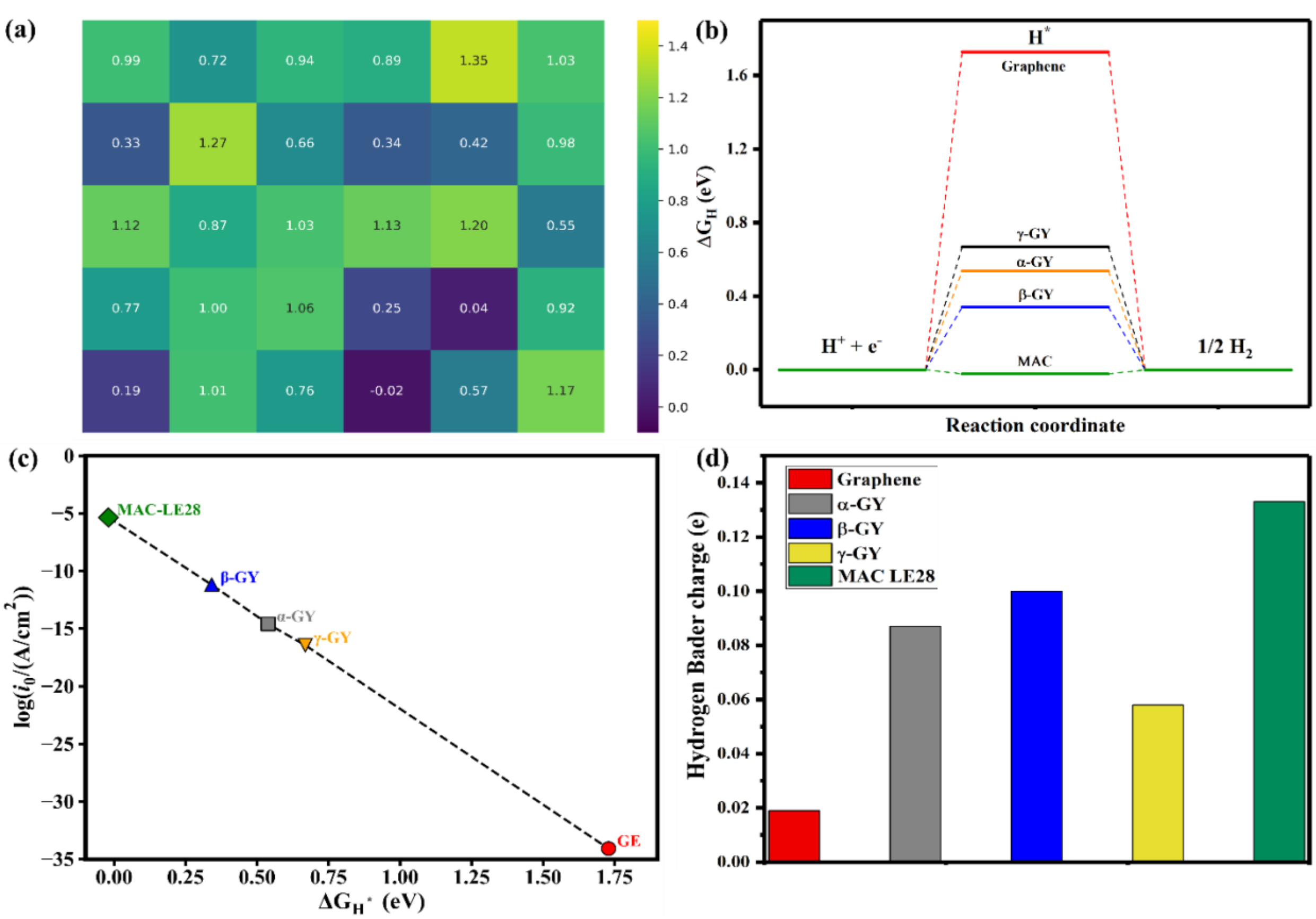

**Figure 4:** **(a)** A heatmap showing the distribution of $\Delta G_H$ of MAC, **(b)** A comparative plot showing the best $\Delta G_H$ values of both crystalline and amorphous structures, **(c)** The exchange current vs $\Delta G_H$ for each material and **(d)** The Bader charge on adsorbed hydrogen on each surface. The term MAC LE28 corresponds to MAC local environment number 28.

The Bader charge on adsorbed hydrogen was calculated to find the extent of adsorption or bonding between hydrogen and the active site, the plot is shown in Figure 4(d). Graphene weakly bonds with the incoming hydrogen ion with a minimal Bader charge of 0.018e. Graphynes showed better adsorption than graphene, with slightly higher charge; however, MAC was higher at 0.14 e. It shows an optimal bond between the hydrogen ion and the catalyst, neither too strong nor too weak.

The total density of states (TDOS) for all species before and after H adsorption is shown in Figure S8. Graphene exhibits semi-metallic zero-bandgap electronic states, whereas β- and γ-GY show small band gaps with empty states near the Fermi level. In comparison, γ-GY possesses a wider bandgap (or more empty states near the Fermi level). α-GY shows semi-metallic behavior like graphene. The local environment of MAC is metallic, as the states are spread across the valence and conduction bands. The adsorption of hydrogen on the surface introduces new states near the Fermi level due to hybridization between H 1s and C 2p orbitals. Furthermore, the projected density of states (PDOS) for all adsorbed surfaces was studied. The plots shown in Figure 5(a)-(e) are the PDOS plots of adsorbed hydrogen and bonded carbon. H-adsorbed graphene showed a sharp hydrogen peak at the Fermi level, but it did not completely overlap with the carbon peaks, indicating a weak interaction. Pristine graphene exhibits a Dirac cone, having no peaks at the Fermi level. But the adsorption of H breaks their $sp^2$ π network, pushing new states at the Fermi level. Graphyne structures exhibit very different PDOS intensities and positions from each other, reflecting differences in hybridization, bonding, and structure. α, β, and γ-GY showed a minor H 1s peak at the Fermi level with negligible interaction with the C 2p orbital. Less intense H peaks in PDOS indicate smaller charge redistribution after H adsorption, possibly due to reduced C-H bond lengths, as shown in Table S1. It could also indicate more stable adsorption sites for graphynes. MAC has a smeared PDOS plot, which agrees with their long-range non-periodicity. MAC LE28 shows a higher number of smaller H peaks in the vicinity of the Fermi level.

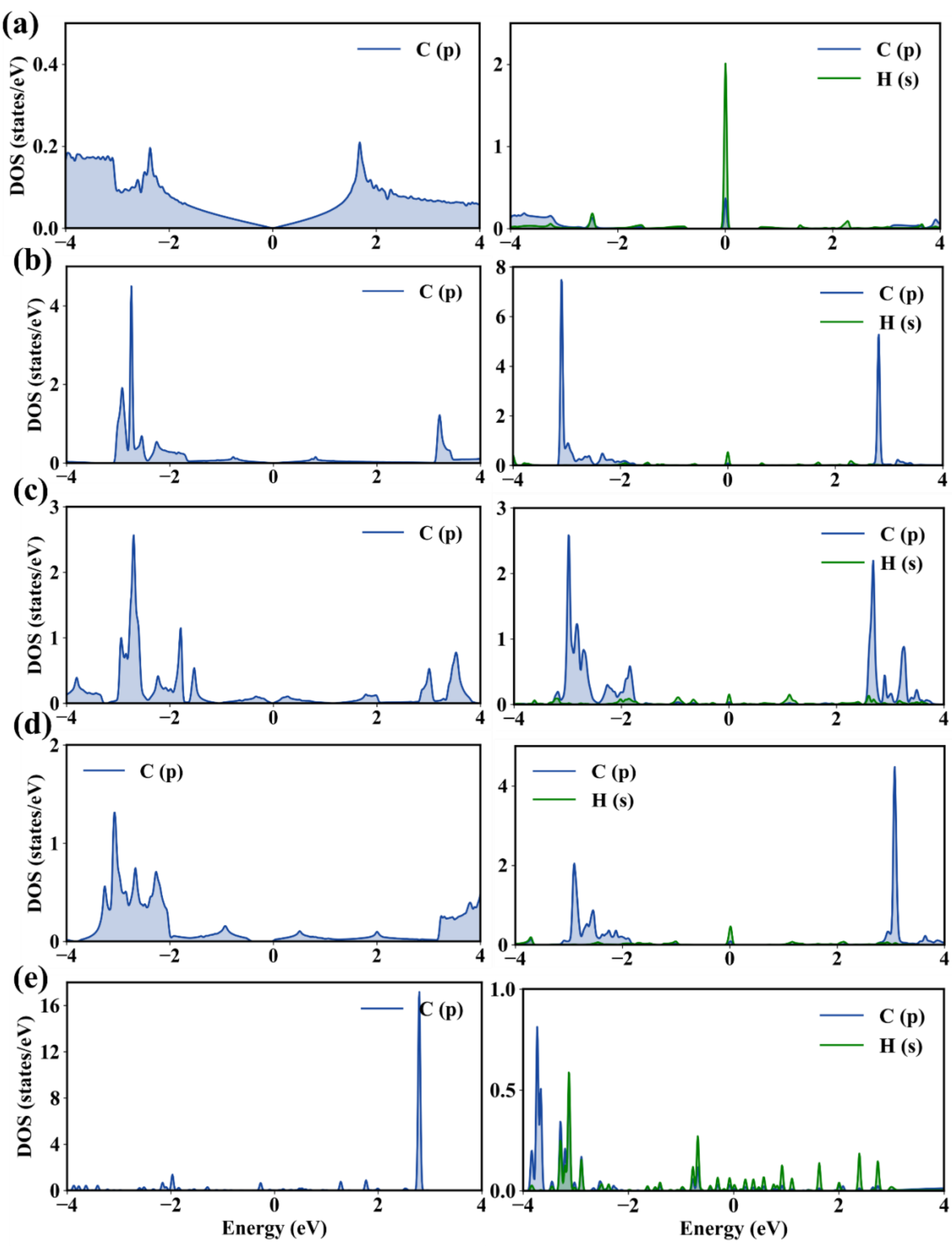

**Figure 5:** Projected DOS analysis of the s orbital of adsorbed hydrogen and the p orbital of carbon attached to the hydrogen. **(a)** α, **(b)** β, **(c)** γ-GY, and **(d)** MAC LE 28

The isolated local environments of MAC with the most favorable $\Delta G_H$ values are shown in Figure 6. Each environment is distinct, and the combination of distinct coordination environments leads to distinct catalytic activities. No direct interpretation of ring distribution and other geometric parameters in relation to activity was found. However, a denser sampling of hydrogen adsorption sites across the surface may reveal any underlying relationships.

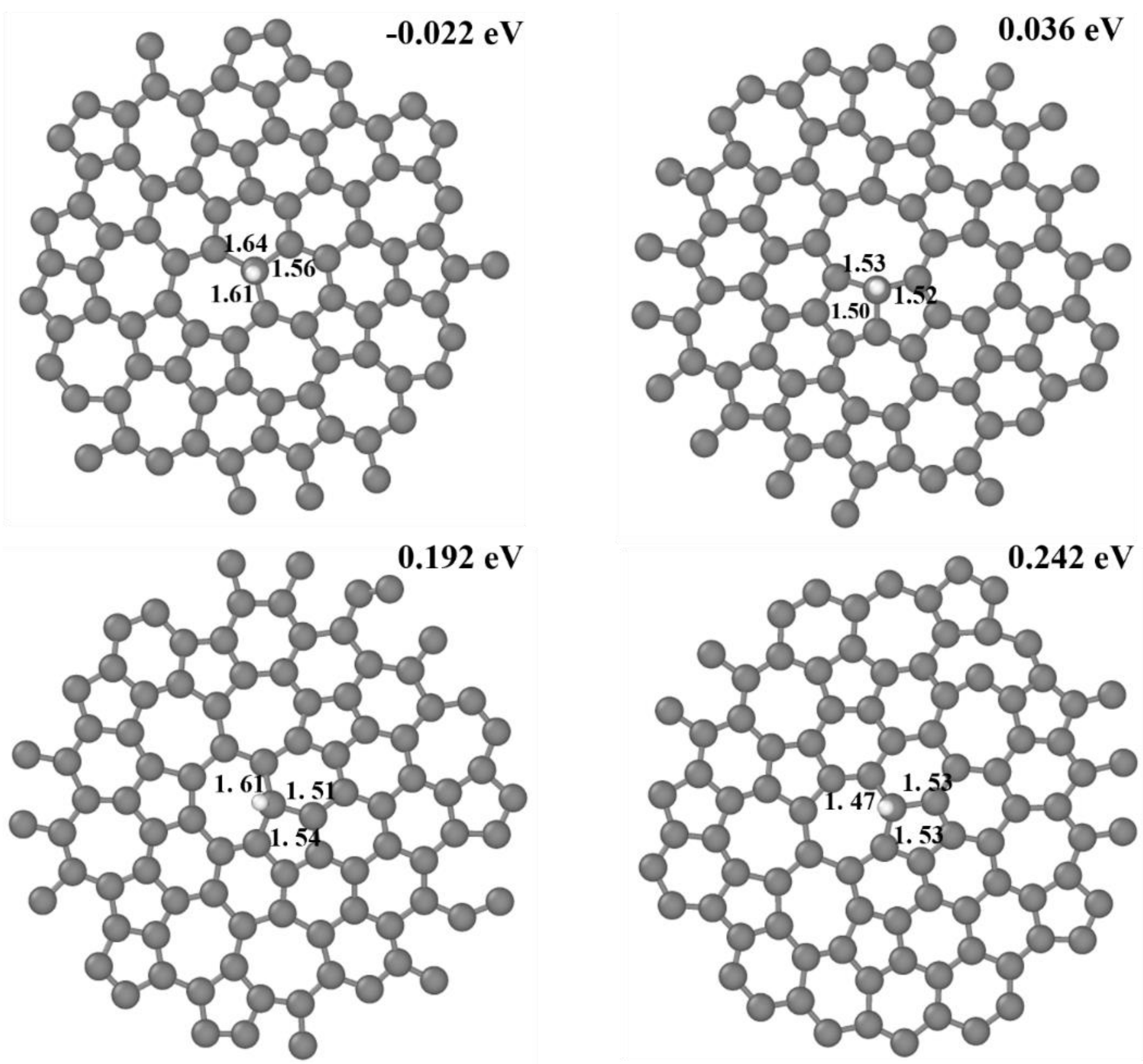

**Figure 6:** Top four active sites of MAC among 30 sampled sites, free energy change, and bond lengths between neighboring atoms to the carbon attached to hydrogen are given. Bond lengths are in Angstroms (Å).

### 3.3.2 Approach 2: MLIP for Hydrogen Adsorption On MAC

The random arrangement of 5, 6, and 7-membered rings in MAC results in a non-uniform free energy landscape. Unraveling how structural variations in the local environment relate to changes in Gibbs free energy would provide clear design principles for optimizing MAC synthesis toward HER applications, which first requires a broader sampling of H adsorption sites. However, evaluating many such sites with DFT is computationally expensive. MLIPs approximate the potential energy surface based on the accurate DFT reference data, and the quality of this approximation governs their performance for a given task. We fine-tune the existing MACE foundation model to predict the Gibbs free energy change of hydrogen adsorption on MAC, using a relatively small dataset. By evaluating $\Delta G_H$ across the MAC surface, we identify a statistical correlation between a set of local structural features and the free energy change. The overall workflow is summarized in Figure 7.

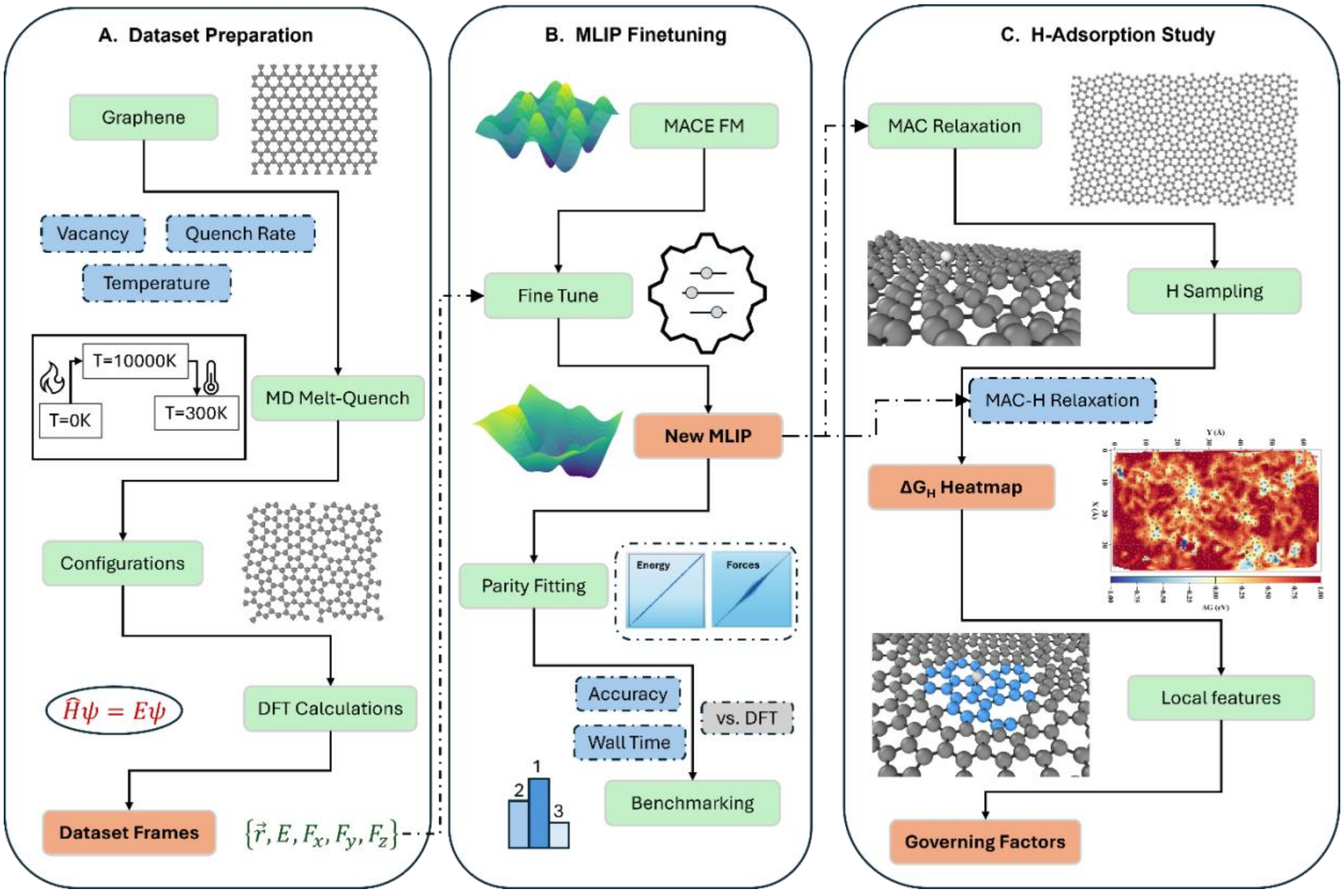

**Figure 7:** Schematic representation of the MLIP workflow for predicting the free energy change landscape of H adsorption and governing structural factors.

### Fine-tuned MLIP fitting

Primarily, the fine-tuned model was tested on 5% random frames of initial training set. The energy and forces parity fittings of the model are shown in Figure 8(a) and (b) respectively. The energy RMSE reached around 1.65 meV/atom, similarly, force RMSE reached to lower value of 29.15 meV/Å. To evaluate the ability of the model to distinguish closely spaced frames, we have tested the model against consecutive frames from a DFT relaxation (ionic cycles of a DFT structure relaxation) outside the dataset. The energy RMSE came out 0.55 meV/atom for static calculation of relaxation frames, and the fit is shown in Figure 8(c). The model was further tested against true relaxation pathways and final energy from DFT. This method of MLIP testing is very important specifically for the problem of our interest. We used structural optimizers in-built inside ASE module to relax the structure using MACE MLIP. Before testing on multiple DFT relaxations, a thorough benchmarking was done against different ASE optimizers. The benchmarking of FIRE and hybrid FIRE+LBFGS optimizers are shown in Figure S9. FIRE is a damped molecular dynamics-inspired optimizer that excels in noisy or rugged energy landscapes like in amorphous systems. It is generally slower near convergence and introduces unnecessary overhead when the potential is close to well-behave. LBFGS is a quasi-Newton method approximating the Hessian with limited memory for efficiency. It excels in converging to near minimum for smooth landscapes but fails for noisy landscapes like in amorphous system. Thus, we combined the optimizers in such a way that the initial rugged landscape will be minimized using early FIRE method until the forces on the system reduce to a threshold of 0.1 eV/Å, followed by final refinement using LBFGS to a force convergence of 0.01 eV/Å. Both optimizers were performing equally well in terms of final energy prediction, although the hybrid FIRE+LBFGS optimizer was chosen for all further calculations. We have tested the MLIP relaxation with 7 to 8 DFT relaxations, including 200 atoms MAC and H adsorbed at the same MAC sites. The plots of energy convergence and force optimization along with wall time (time cost) of MLIP and DFT relaxations are shown in Figure S10. The total energy values after H adsorption on multiple sites with both the relaxation methods are summarized in Table S2. A correlation map of individual atomic displacements from initial (pre-relaxation) to final (post-relaxation) configurations between MLIP and DFT is shown in Figure 8(d). No exceptional or outlier atomic displacements were observed during MLIP relaxation, and we achieved a strong correlation ($R^2 = 0.97$). This indicates that the MLIP closely mirrors DFT's relaxation pathways.

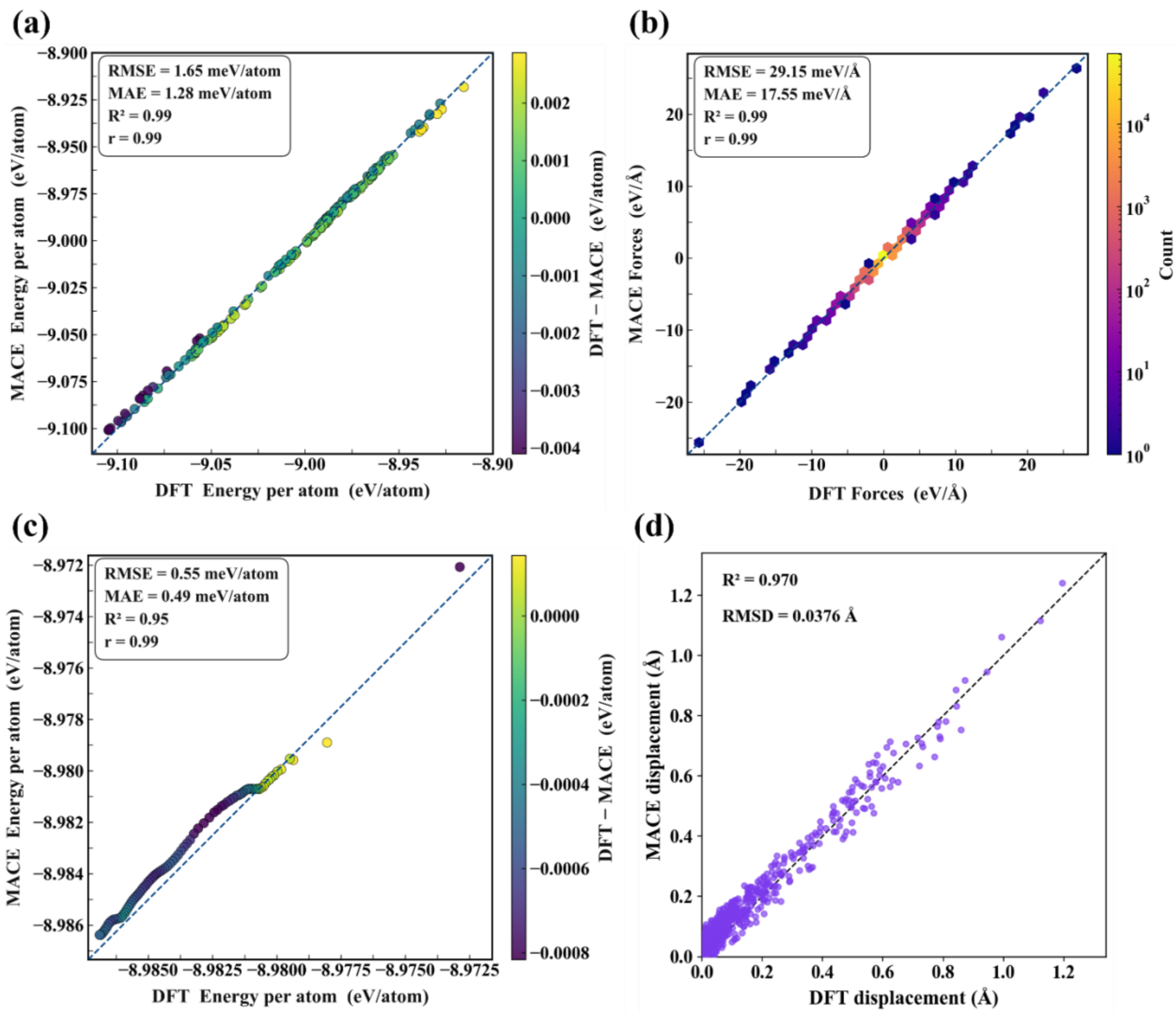

**Figure 8: (a)** Energy and **(b)** force predictions of the fine-tuned model against the test dataset, **(c)** energy fitting across individual frames from a DFT structural relaxation trajectory, and **(d)** a correlation plot showing per-atom displacements from pre-relaxation to post-relaxation structures, comparing MLIP and DFT relaxation patterns.

To rely on the MLIP for predicting Gibbs free energy of H adsorption on MAC, near-ideal agreement between MLIP and DFT final relaxed energies is essential. The Gibbs free energy of H adsorption ($\Delta G_H$) is accurately determined from the final energies of the relaxed bare MAC and H-adsorbed MAC structures as per Equation 1. Inaccurate predictions of the final energy for H adsorbed at a specific MAC site may yield wrong $\Delta G_H$ signature for that site. Conversely, errors in the bare MAC reference energy would propagate to misestimate $\Delta G_H$ across all adsorption sites.

This error propagation throughout structural relaxation was rigorously analyzed using the aforementioned DFT relaxation energy calculation and displacement fitting.

As the next step, we sampled single H adsorption on random top and bridge sites of ~900 atoms MAC system shown in Figure 9(a), totaling around 1183 sites. Prior to sampling, MLIP-based structural relaxation was performed on the MAC surface. Relaxed MAC adopts a complex buckled configuration, positioning most carbon atoms beyond the initial plane as shown in Figure S11. It exhibits a ripple-like morphology across the surface. Each H-adsorbed MAC configuration underwent individual MLIP relaxation to obtain site-specific adsorption energies. $\Delta G_H$ values were evaluated for ~1183 sites across the MAC surface, yielding the heatmap shown in Figure 9(b).

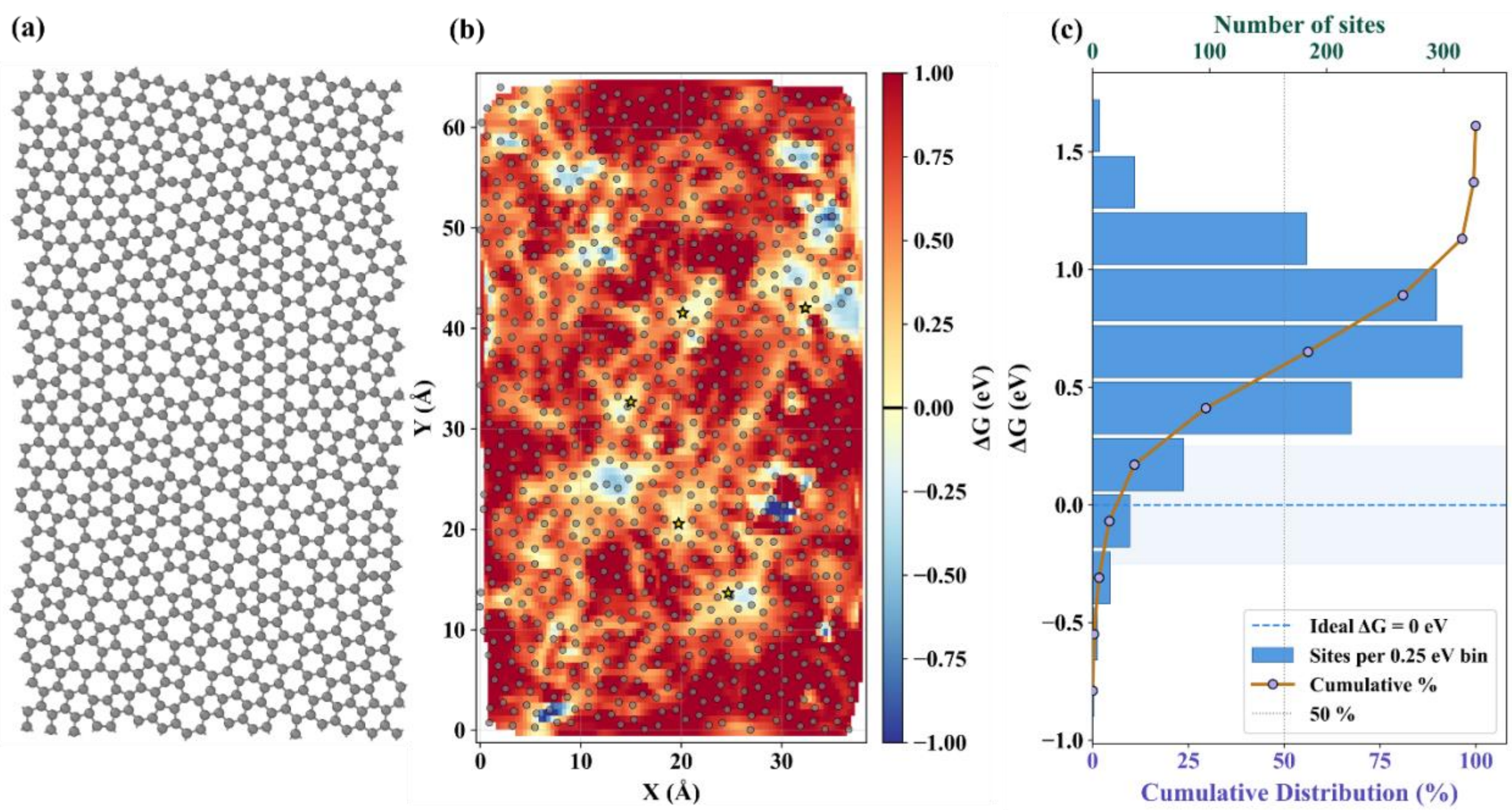

**Figure 9: (a)** MAC structure after MACE MLIP relaxation, **(b)** $\Delta G_H$ heatmap of MAC surface, **(c)** $\Delta G_H$ distribution and cumulative analysis.

The light-yellow region in the heatmap shows preferable sites for H adsorption ( $\Delta G_H \approx 0\ eV$). The red region rejects H adsorption, where the blue region traps the adsorbed H. Consistent with DFT local environment analysis from approach 1, the MAC surface exhibits a highly site-sensitive $\Delta G_H$ landscape, with free energy changes spanning -0.91 eV to +1.7 eV. The distribution of sites across different $\Delta G_H$ intervals along with a cumulative statistic is provided in Figure 9(c). Around 60% of sites have $\Delta G_H$ value less than +0.75 eV and around 15% sites have $\Delta G_H$ value less than

+0.25 eV. This is a promising landscape compared to pristine graphene ($\Delta G_H = +1.73$ eV) and other graphynes.

We examined the local environment around each individual site to better understand the key factors influencing optimal H adsorption. Structural features of the immediate local environment (within the 5Å spherical shell) were extracted and estimated to have a simple statistical correlation against the $\Delta G_H$ at that site. The height deviation of each carbon atom from the mean $z = 0$ surface is visualized via a color gradient in Figure 10(a). A more specific depiction of XZ plane of MAC along with the adsorbed hydrogen, carbon neighbor of hydrogen and atoms in the local environment is shown in Figure 10(b). We extracted a total of 13 local features for each individual site. Those features, their explanation and significance are summarized in Table 3. Also, a plot showing the ring size distribution in the local environment is shown in Figure 10(c), the smallest ring is highlighted in the same plot.

| **Feature** | **Explanation and Significance** |
| --- | --- |
| Coordination Number | Coordination number of the nearest C atom to H represents hybridization. |
| Mean angle | Mean bond angle around the adsorption-site C, deviations represent strain and curved surfaces. |
| Angle std | Standard deviation of bond angles. High values indicate geometric disorder in bonding environments. |
| Nearest C mean height | Z-displacement of the adsorption C above the plane of its neighbors. Non-zero values indicate out-of-plane puckering. |
| Local density | Number of atoms within 5 Å of adsorption C, represents how "crowded" the site is. |
| Curvature | Normalized spread of first-shell bond lengths. High curvature indicates a non-flat surface. |
| Bond variation | Variance of first-shell bond lengths. Uniform bonds give low variance; stretched bonds give high variance, indicating local strain |
| Hexagonal order | Bond-orientational order. Value near 1implies perfect hexagonal order (graphene), near 0 implies amorphous. |
| Neighbor CN variation | Variance of coordination numbers of the nearest neighbors. High variance means the site is near to an extremely asymmetric region. |
| 5-membred rings | Number of 5 carbon atom rings, increases local density or vice versa. |
| 6-membred rings | Number of 6 carbon atom rings. |
| 7-membred rings | Number of 7 carbon atom rings, decreases local density or vice versa. |
| Smallest ring | Size of the smallest ring containing the adsorption site. |

**Table 3**: Examined local features, their explanations, and importance.

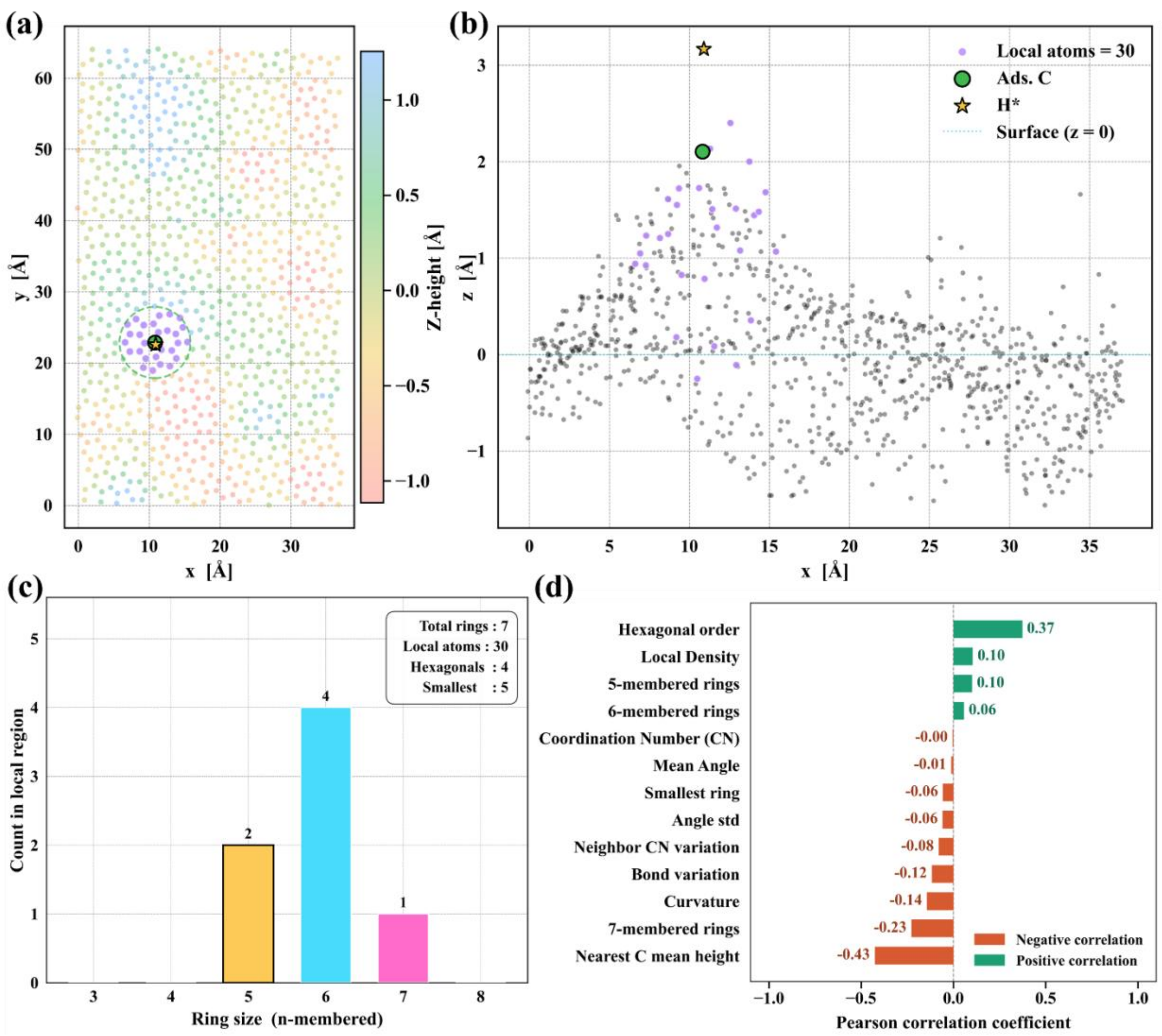

**Figure 10:** Height variation of carbon atoms from the perspective of **(a)** XY plane and **(b)** XZ plane. Local environment of 5Å radius is shown by green dotted circles. **(c)** The ring size distribution and **(d)** correlation map between local features and $\Delta G_H$.

Using the local environment data extracted for each site, we calculated Pearson correlation coefficients between each feature and $\Delta G_H$. The correlation coefficient indicates the positive or negative relationship between two properties, a positive value shows they change in the same direction, while a negative value shows one decreases as the other increases. The MAC surface exhibited an average $\Delta G_H$ of +0.65 eV overall, so features with negative correlation coefficients would lower $\Delta G_H$, improving HER performance. Sites retaining more graphene-like angular distribution (hexagonal order) and 6-membered rings show reduced catalytic efficiency. A higher proportion of 5- and 6-membered rings elevates local density and thus increases $\Delta G_H$. The negative correlation of 7-membered rings with $\Delta G_H$ possibly arises because they reduce local density (bigger voids), which lowers $\Delta G_H$ and enhances HER performance. Bond variation negatively correlated to $\Delta G_H$. An increase in bond length variation raises the local strain, which in turn strengthens hydrogen binding [52]. As the curvature of the surface and height (relative to average plane of neighbors) of the carbon atom bonded to hydrogen increases, $\Delta G_H$ reduces at that site. This link between local curvature/height and catalytic performance is well-documented in prior experimental and theoretical research. Notably, the surprising strong catalytic activity of nano-rippled graphene for $H_2$ dissociation was attributed to surface corrugations, with hydrogen splitting becoming thermodynamically favorable only when curvature surpassed ~10% [53]. Other studies showed that the carbons on curved surfaces exhibit increased chemical reactivity due to diminished electronic delocalization and higher σ-bond character [54], [55]. From all the feature observations, activity is not determined by a single feature alone but rather requires a combined feature matrix. Finetuning this feature matrix is essential to develop a fully functional MAC catalyst capable of improved HER performance. Table S3 provides a thorough summary of $\Delta G_H$ values for established crystalline and amorphous HER catalysts. Fine-tuning the feature matrix to synthesize MAC could yield a highly effective catalyst rivaling current benchmarks.

## Conclusions

We have studied the HER performance of 30 randomly selected sites in Monolayer Amorphous Carbon (MAC) and compared it with that of crystalline materials such as graphene and different types of graphyne using Density functional theory. The MAC prepared via melt-quench simulation showed a diverse ring distribution of 5, 6, and 7-membered rings, as well as a hybridization distribution of $sp^2$ and $sp^3$ carbons. Hydrogen adsorption on MAC was carried out individually on an isolated local environment of 9.0 Å. MAC showed an impressive Gibbs free energy change value ($\Delta G_H$) of -0.02 eV for a particular site. The $\Delta G_H$ values for all 30 sites ranged from -0.02 eV to +1.38 eV, which is better than those of pristine graphene. Furthermore, Bader charge analysis of adsorbed hydrogen shows optimal adsorption on the MAC site and on β-graphyne, but weaker adsorption on other surfaces. The adsorption study was further extended to produce a full $\Delta G_H$ heatmap of MAC using finetuned MACE foundation model. The model showed an energy and force RMSE of 1.65 meV/atom and 29.15 meV/Å respectively. After confirming the model's optimal performance on DFT relaxation pathways, a single hydrogen atom was sampled at approximately 1183 different sites on MAC. The $\Delta G_H$ values varied from -0.91 eV to +1.70 eV. Around 15% of sites on MAC showed $\Delta G_H$ values less than +0.25 eV, a range typically observed only with noble-metal-based catalysts. Roughly 60% of the sites exhibited $\Delta G_H$ values below +0.75 eV, significantly outperforming pristine graphene. From the local feature correlation study, it was observed that a planar graphene-like environment and higher local density reduce HER efficiency, while a more loosely packed environment arising from 7-membered rings and surface ripples promotes HER activity. As a next step, a well-defined mapping between structural diversity and experimentally feasible synthesis methods is to be developed. By combining these two distinct mappings, theory can predict the most promising synthesis routes for catalytically active MACs for hydrogen evolution, and potentially, a host of other electrochemical reactions.

## Acknowledgements

The authors acknowledge IIT Gandhinagar's PARAM Ananta supercomputing facility for conducting all the simulations reported in this work.

## Supplementary Information

## Harnessing Structural Disorder: Unraveling Hydrogen Evolution in Monolayer Amorphous Carbon via First-Principles Simulations and Machine-Learned Potentials

Sreehari M S, Ashutosh Krishna Amaram, Raghavan Ranganathan*

Department of Materials Engineering, Indian Institute of Technology Gandhinagar, Gandhinagar, 382355, Gujarat, India

*Corresponding author: rraghav@iitgn.ac.in

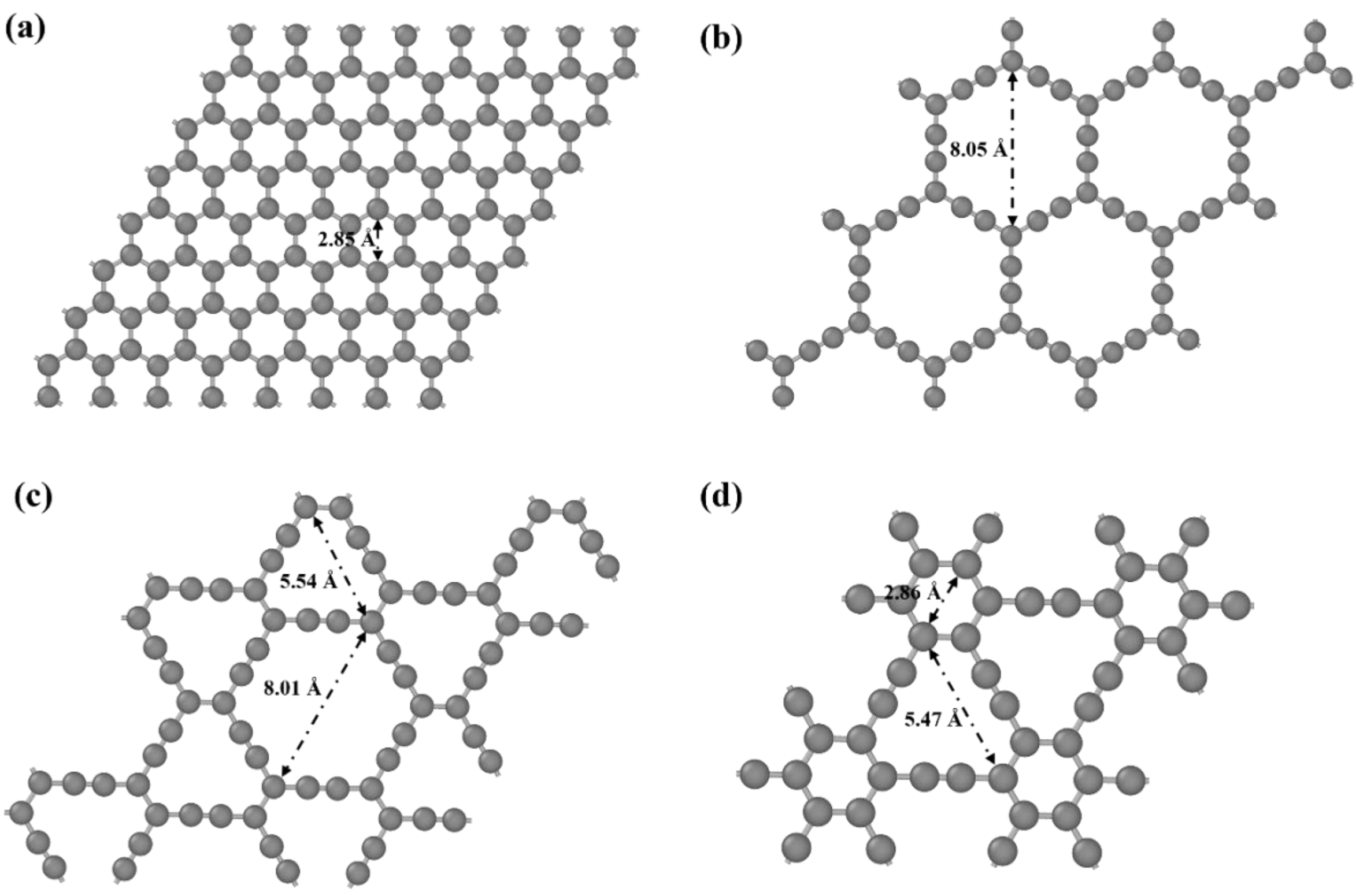

**Figure S1**: The maximum interatomic distance within a ring is depicted for species (a) graphene (b) α-GY, (c) β-GY, (d) γ-GY.

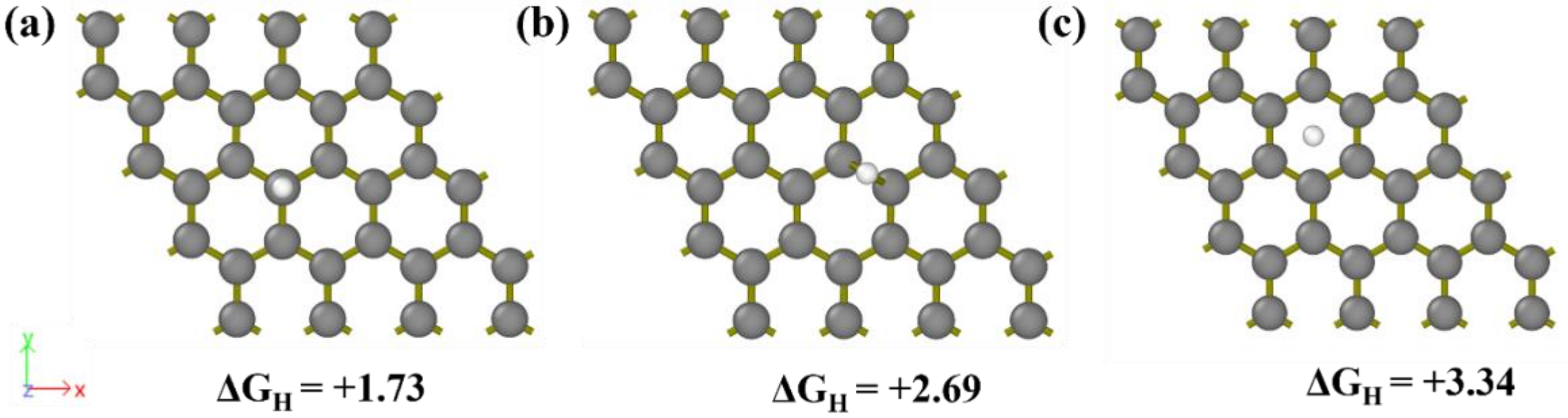

**Figure S2**: The possible sites of H adsorption in graphene (before structure relaxation)

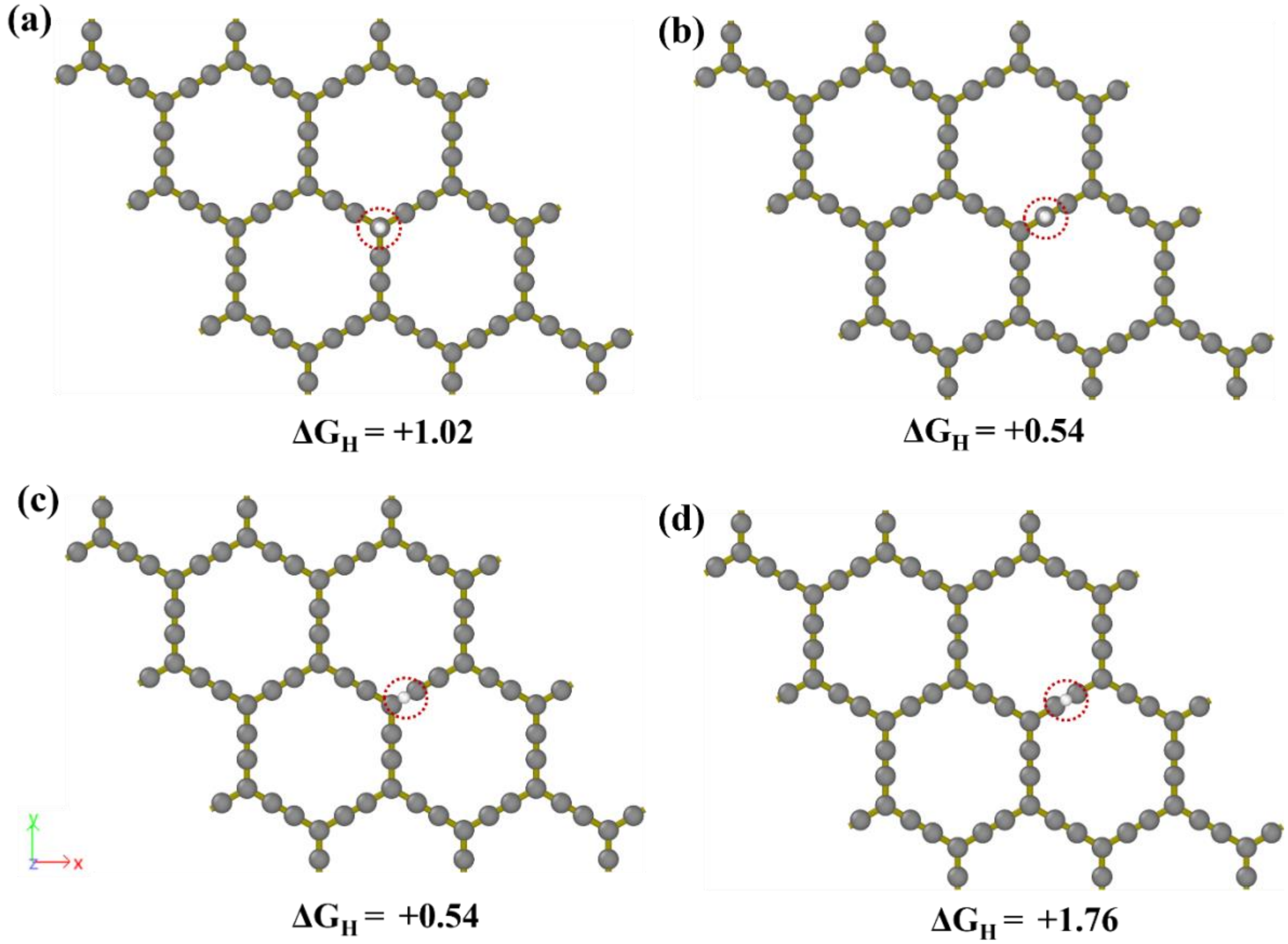

**Figure S3**: The possible sites of H adsorption in α-GY (before structure relaxation)

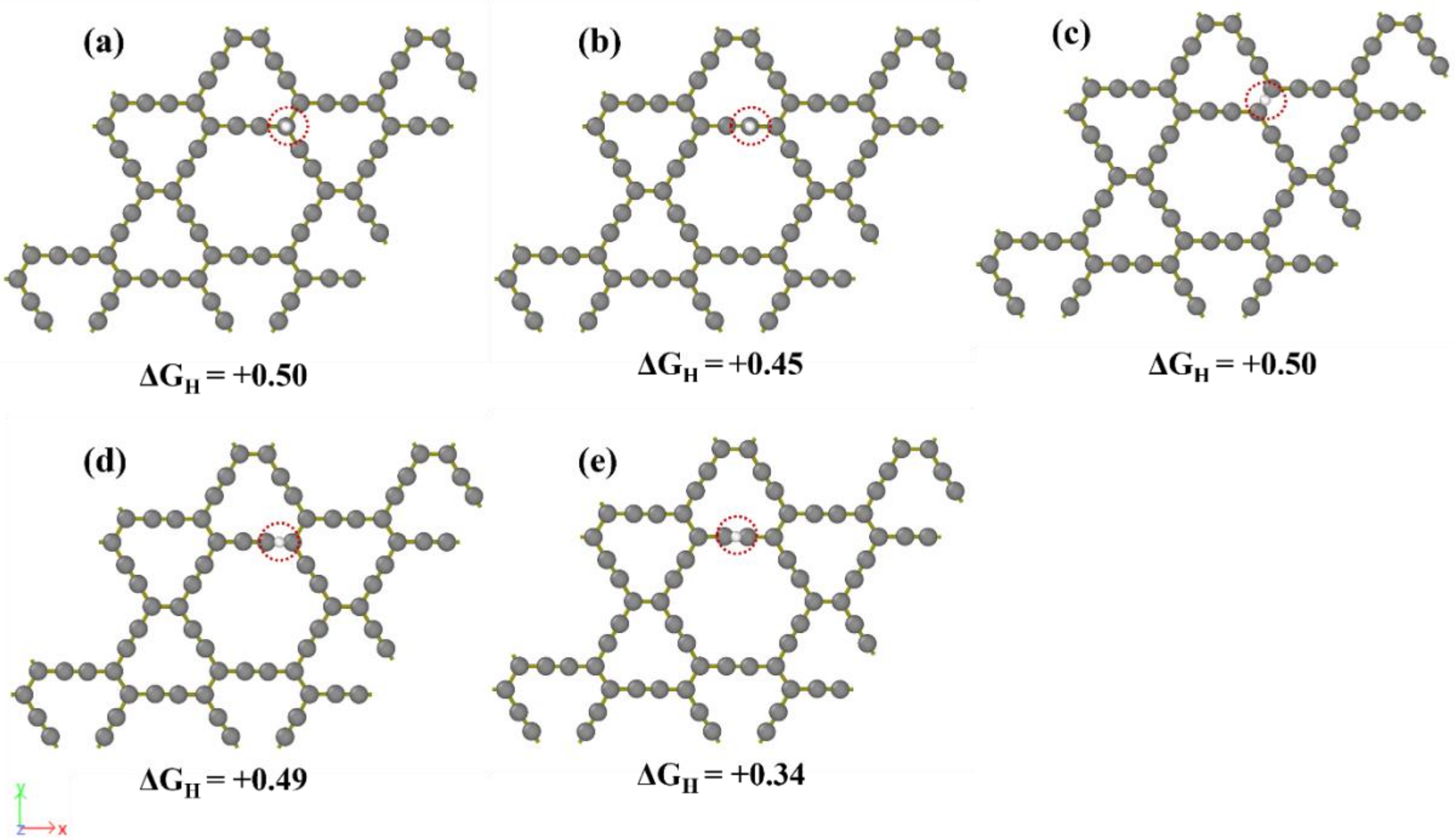

**Figure S4**: The possible sites of H adsorption in β-GY (before structure relaxation)

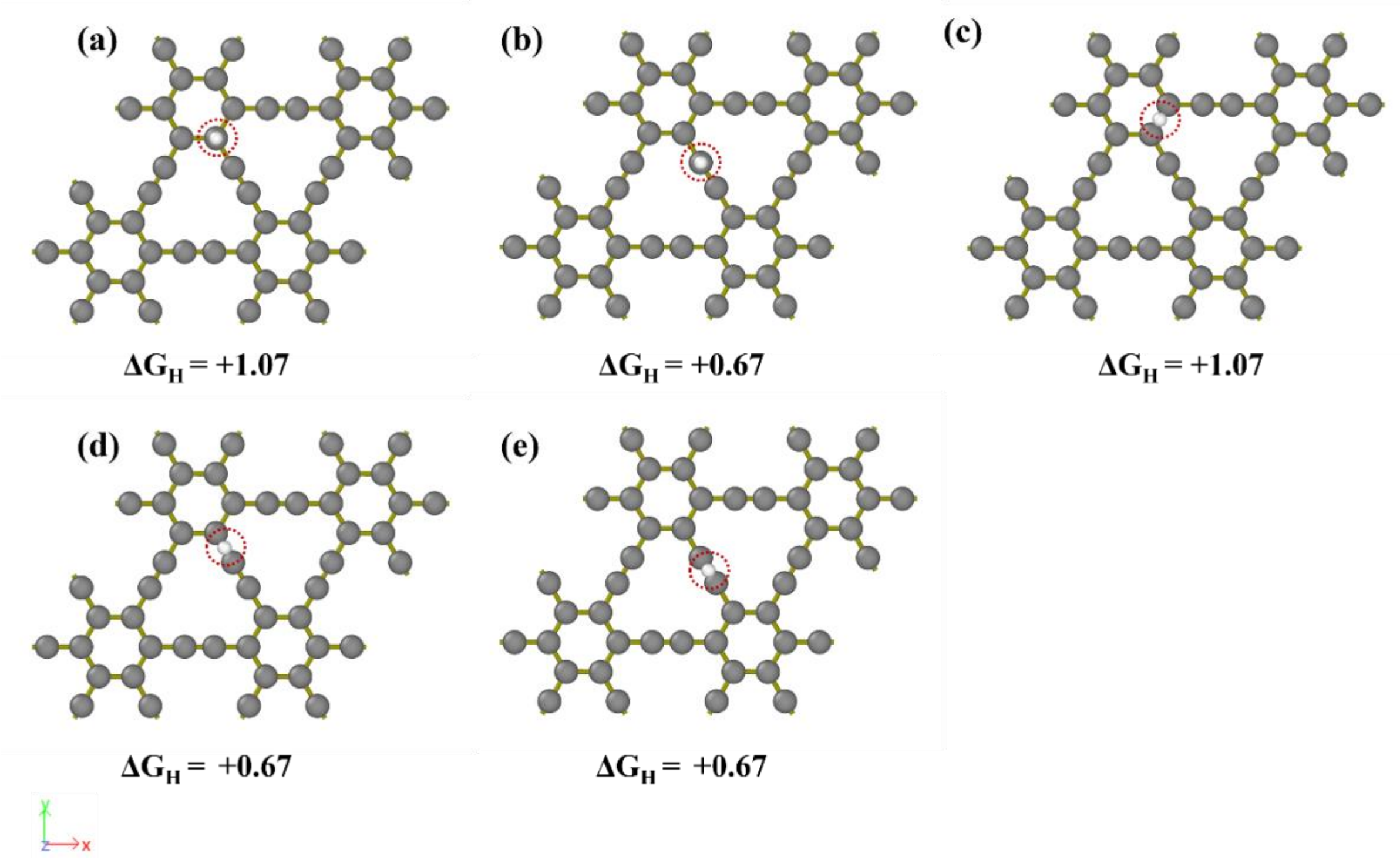

**Figure S5**: The possible sites of H adsorption in γ-GY (before structure relaxation)

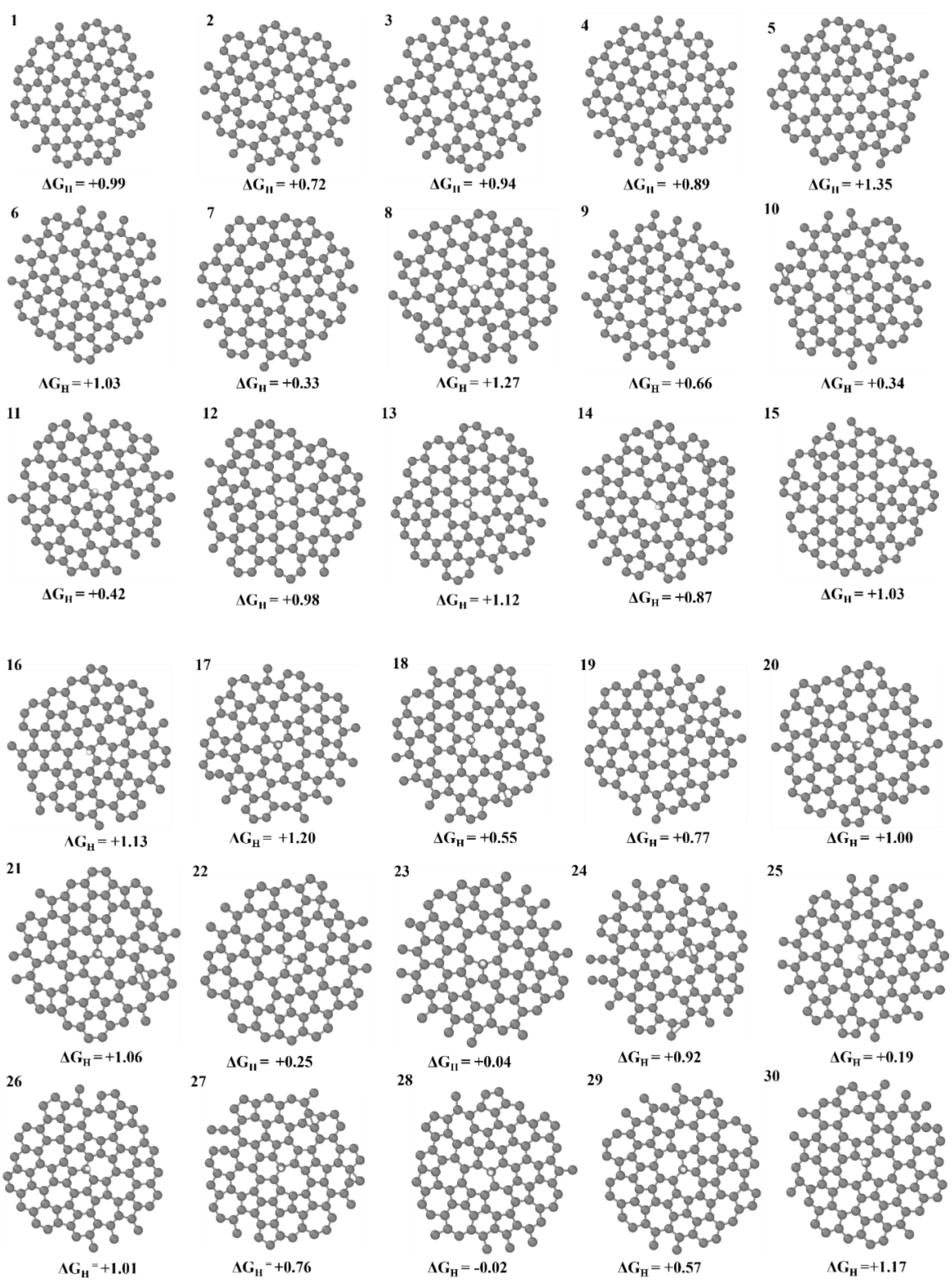

**Figure S6**: Hydrogen adsorbed on 30 random sites of MAC local environment (after structure relaxation)

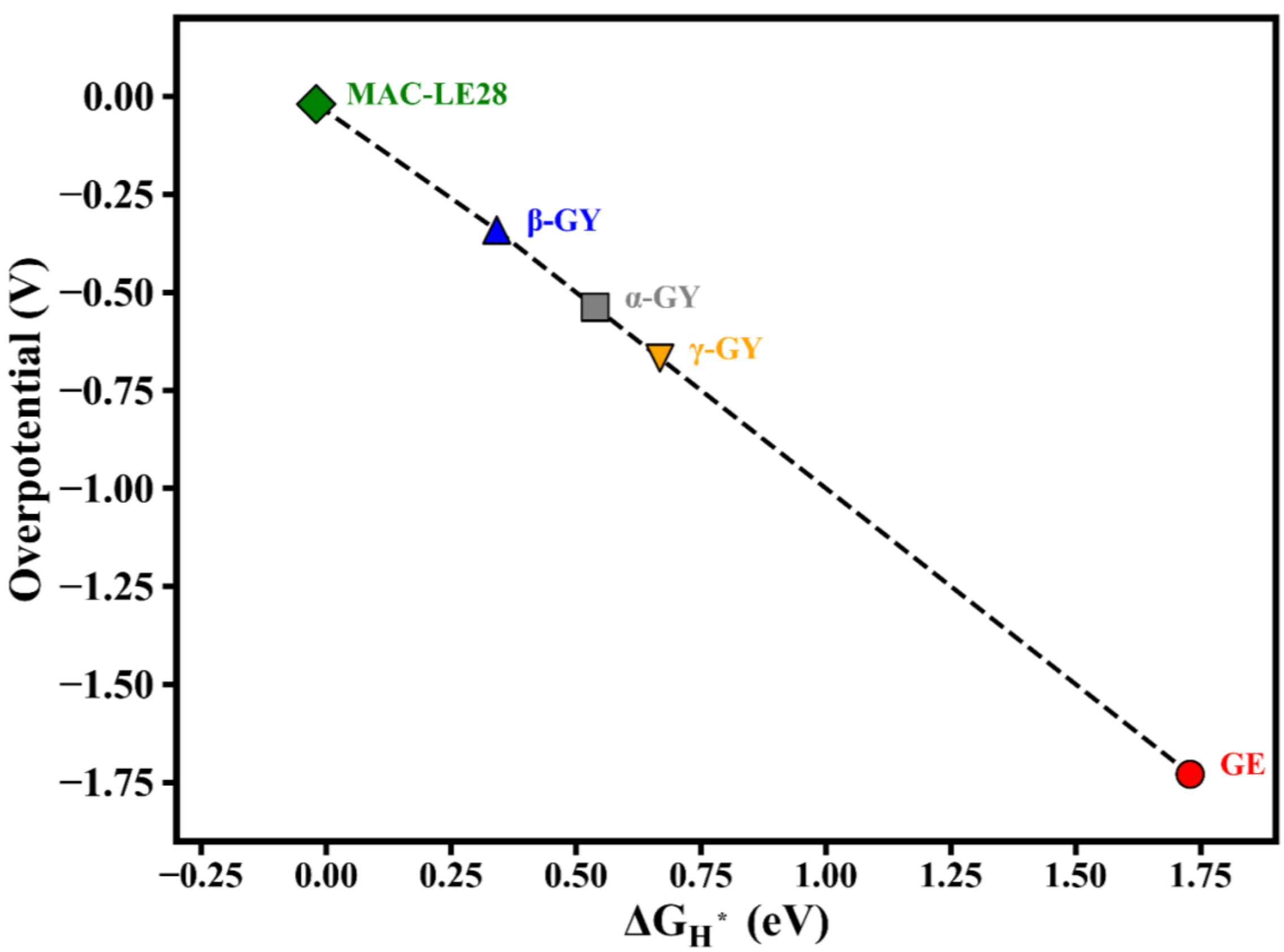

**Figure S7:** Theoretical overpotential vs $\Delta G_H$ for graphene (GE), α, β, γ-GY and MAC LE 28

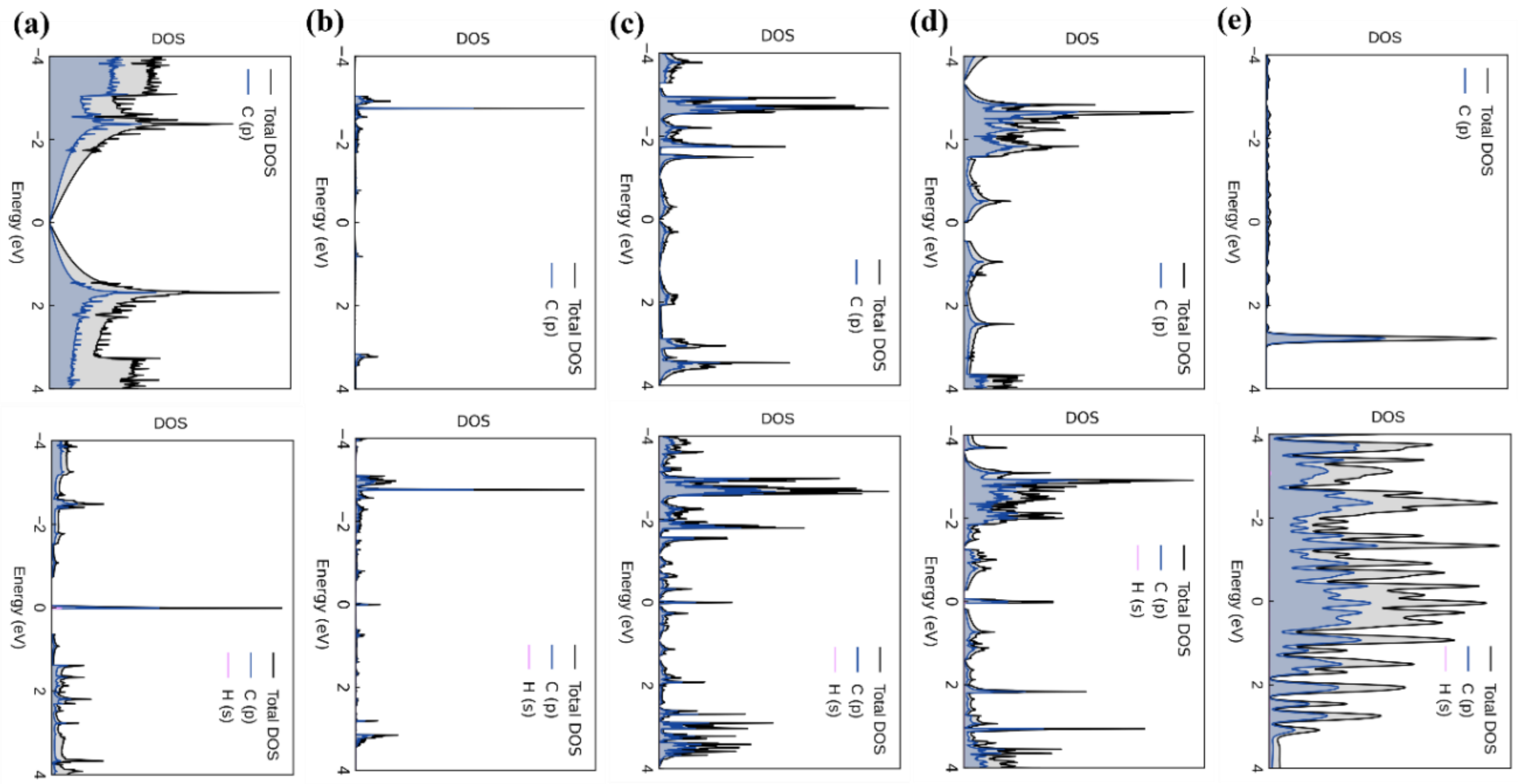

**Figure S8**: Total density of states of (a) graphene, (b) α, (c) β, (d) γ-GY and (e) MAC LE 28 before and after H adsorption.

| **Species** | Graphene | alpha | beta | gamma | MAC LE28 |
|---|---|---|---|---|---|
| **C-H Bond length (Å)** | 1.128 | 1.104 | 1.108 | 1.109 | 1.115 |

**Table S1**: C-H Bondlengths in H adsorbed carbon derivatives

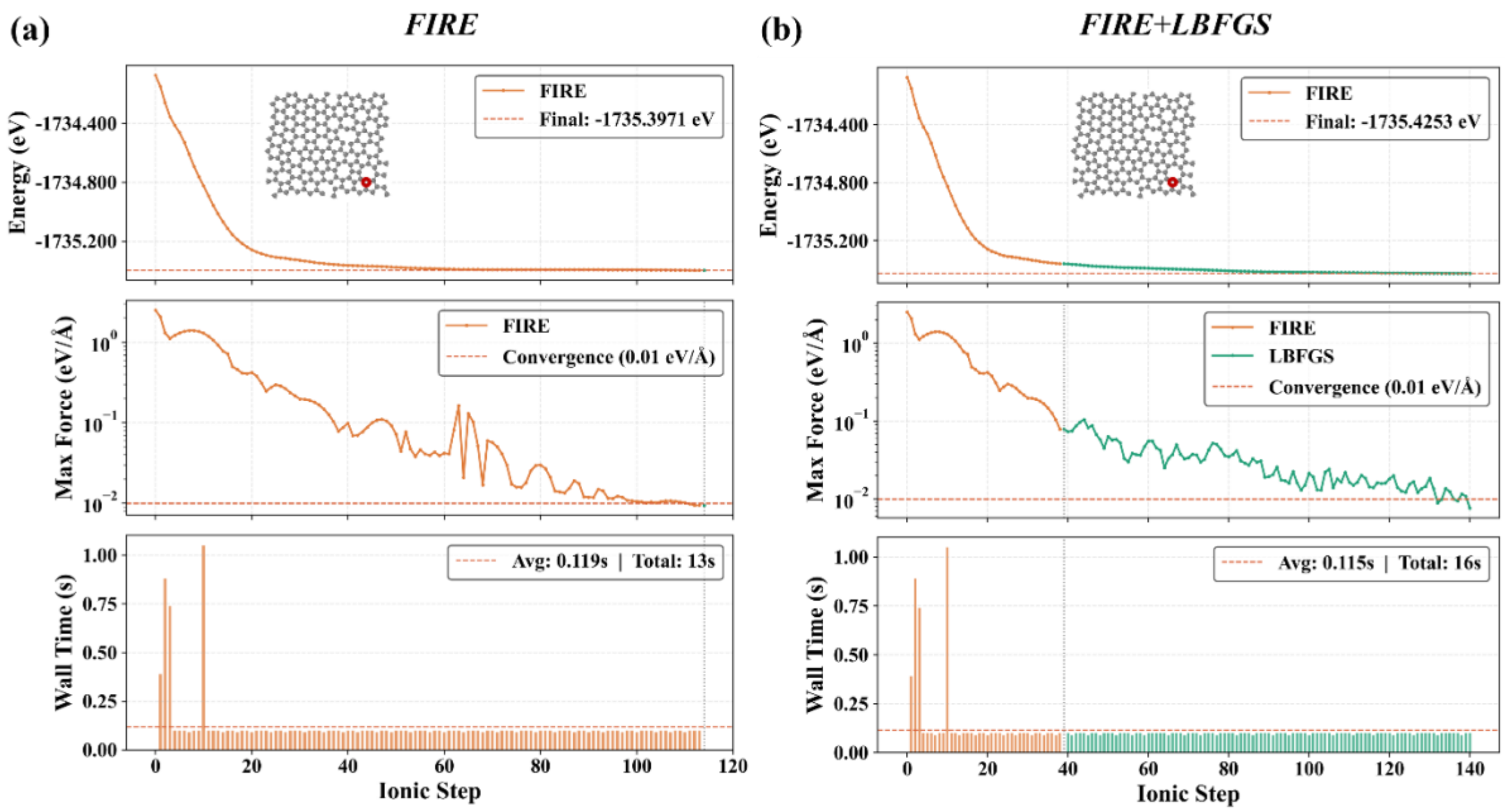

**Figure S9**: Comparison of ASE optimizers for relaxing H adsorption on MAC: **(a)** FIRE versus **(b)** FIRE + LBFGS hybrid. The small red circle on the inline MAC structure marks the H adsorption site.

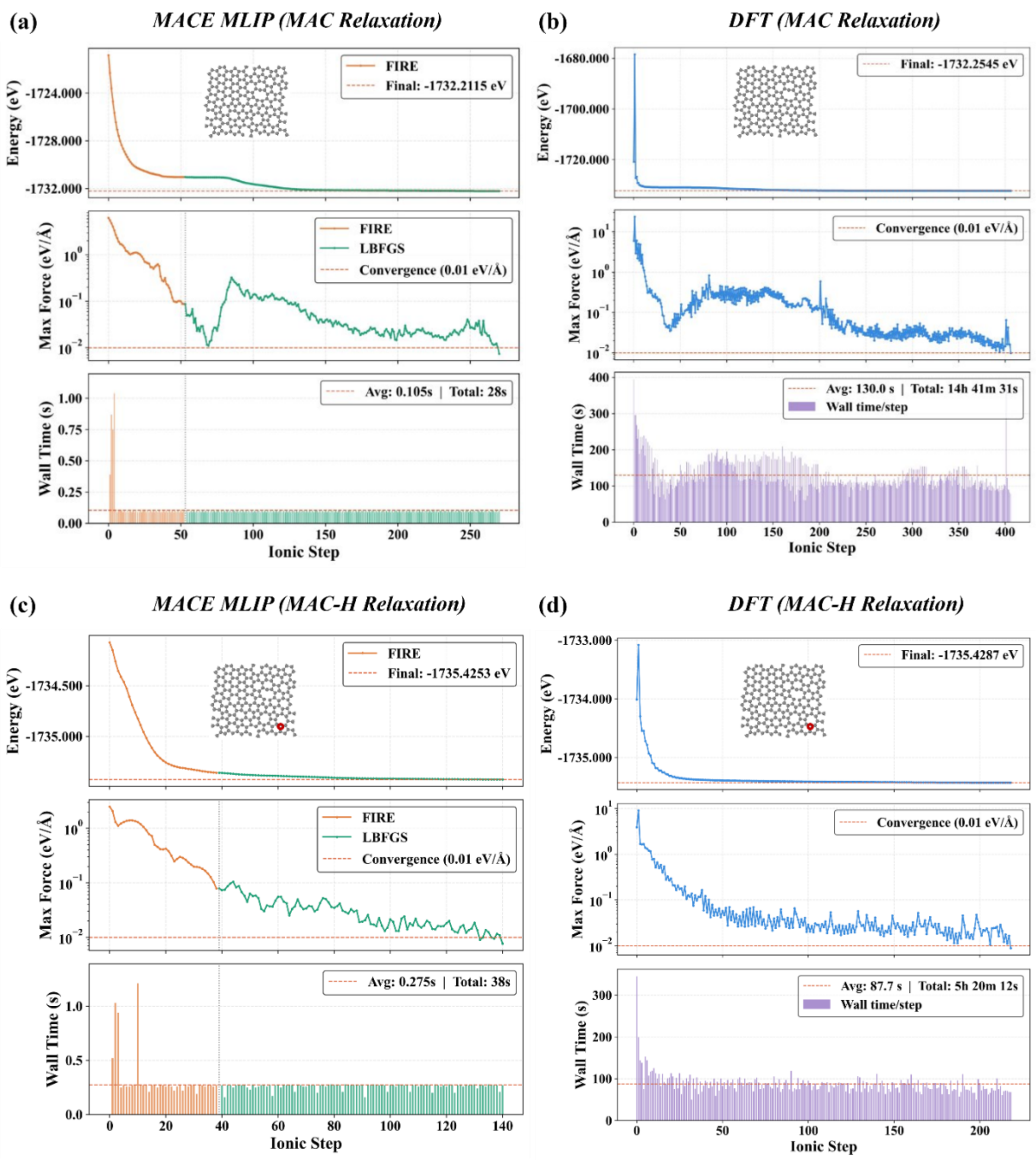

**Figure S10**: Structure relaxation of MAC with **(a)** MACE MLIP and **(b)** DFT, H adsorption relaxation on MAC using **(c)** MACE MLIP and **(d)** DFT.

| Sites | Method | Energy(eV) | Wall Time (s) |
|---|---|---|---|
| 1 | MACE | -1734.8635 | 13 |
| | DFT | -1734.7943 | 8669 |
| 2 | MACE | -1735.4253 | 16 |
| | DFT | -1735.4287 | 19212 |
| 3 | MACE | -1734.8186 | 14 |
| | DFT | -1734.9781 | 11988 |
| 4 | MACE | -1735.2413 | 13 |
| | DFT | -1735.0902 | 10553 |
| 5 | MACE | -1735.0331 | 28 |
| | DFT | -1735.0404 | 9135 |
| 6 | MACE | -1735.1783 | 17 |
| | DFT | -1734.9714 | 18846 |
| 7 | MACE | -1735.3379 | 13 |
| | DFT | -1735.4768 | 12093 |

**Table S2:** Summary of final relaxed energies and computational time costs for MACE MLIP vs. DFT across 7 H adsorption sites. MLIP relaxation was done with GPU, whereas DFT relaxation was done with VASP on CPU.

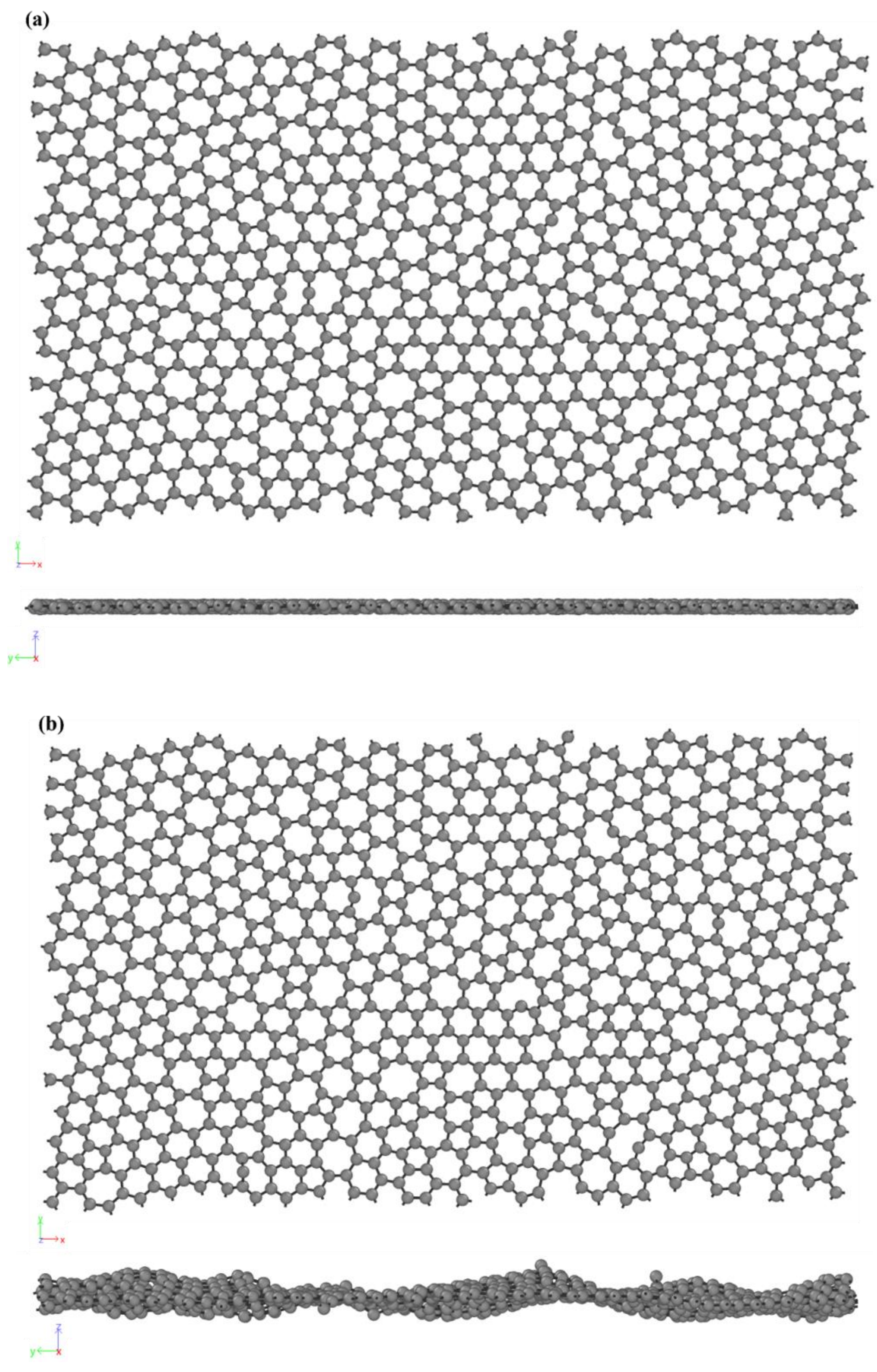

**Figure S11:** The structure of MAC **(a)** before and **(b)** after MACE MLIP relaxation

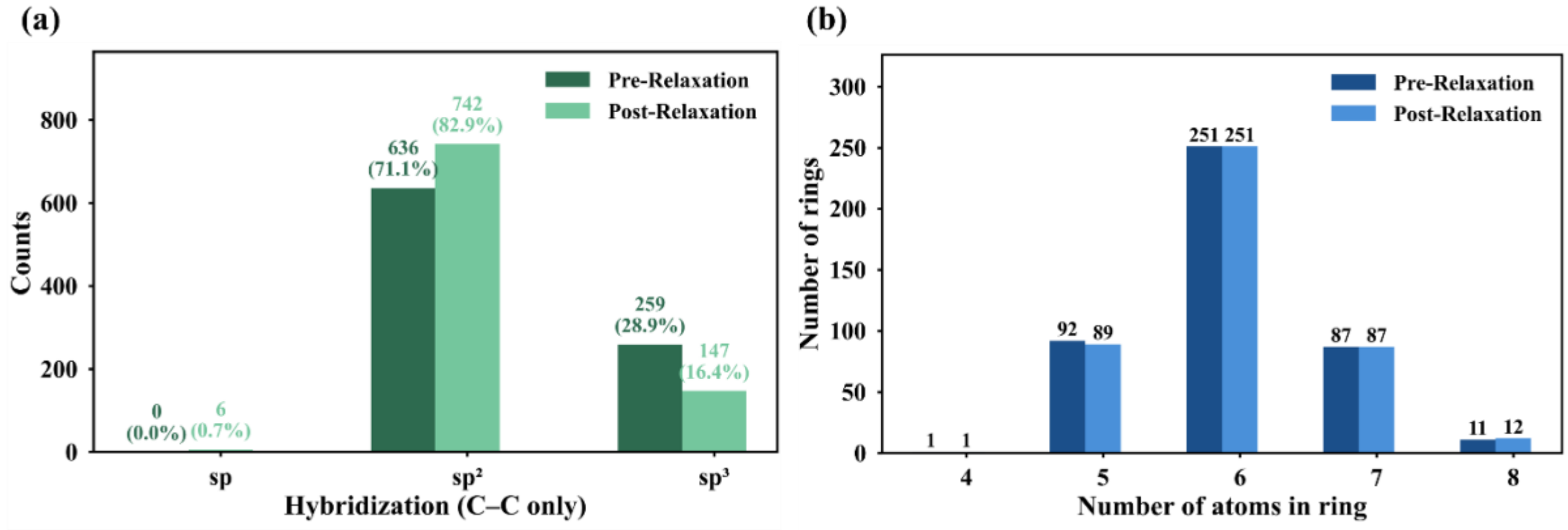

**Figure S12**: **(a)** Hybridization and **(b)** ring size distribution in the 895-atom MAC system before and after MLIP relaxation.

| | Sl. No: | Catalyst | ΔG (eV) | Reference |
|---|---|---|---|---|
| **Crystalline catalysts** | 1 | Pristine graphene (C) | 1.83 | [1] |
| | 2 | N-doped C shell | 1.18 | |
| | 3 | CoMoP | -0.55 | |
| | 4 | CoMoP@pristine-C | 0.43 | |
| | 5 | CoMoP@C (N-dopants) | 0.10 | |
| | 6 | Pt | -0.09 | [2] |
| | 7 | Ni | -0.27 | |
| | 8 | Mo | -0.37 | |
| | 9 | Ni-Mo alloy nanotube | 0.11 (On Ni site)<br>0.35 (On Mo site) | [3] |
| | 10 | $Ni_3Mo$ | -0.18 | [4] |
| | 11 | Ni | -0.30 | [5] |
| | 12 | $Ni_4Mo$ nanoparticle | -0.09 | |
| | 13 | Crystalline $PtSe_2$ | >1.0 | [6] |
| **Amorphous catalysts** | 14 | Defective $PtSe_x$ ($x \approx 1.5$) | 0.5 to 1.0 | |
| | 15 | Amorphous $PtSe_{1.33}$ | -0.2 to 0.2 | |
| | 16 | $Pd_3P_2S_8$ (crystalline) | 1.15 | [7] |
| | 17 | $LiPd_3PS_7$ | -0.09 | |
| | 18 | Monolayer Amorphous Carbon | -0.02 to +1.38 | **This work (from DFT)** |

**Table S3**: Comparison table of this work with previously reported works.